\RequirePackage{fix-cm}
\documentclass[twocolumn,epjc3]{svjour3}  
\smartqed  
\RequirePackage{graphicx}
\RequirePackage{mathptmx}      
\RequirePackage{latexsym}
\RequirePackage[numbers,sort&compress]{natbib}
\RequirePackage[colorlinks,citecolor=blue,urlcolor=blue,linkcolor=blue]{hyperref}
\usepackage{amsmath} 

\usepackage{amssymb}
\usepackage{multirow}
\usepackage{multicol}
\usepackage{graphicx}
\usepackage{amsmath}
\usepackage{lineno}
\usepackage{setspace}
\usepackage{subcaption}
\usepackage{listings}
\usepackage{xspace}
\usepackage{xcolor}
\usepackage{float}
\usepackage{comment}
\usepackage{soul}
\usepackage{wasysym}

\begin{document}

\newcommand{\geant}{\textsc{Geant4}\xspace}
\newcommand{\ncrystal}{\textsc{NCrystal}\xspace}
\newcommand{\toucans}{\textsc{TOUCANS}\xspace}
\newcommand{\tripoli}{{\textsc{Tripoli-4}}\textsuperscript{\textregistered}\xspace}
\newcommand{\mcnp}{{\textsc{MCNP6.2}}\xspace}
\newcommand{\fifrelin}{\textsc{Fifrelin}\xspace}
\newcommand{\fifradina}{\textsc{Fifradina}\xspace}
\newcommand{\iradina}{\textsc{Iradina}\xspace}
\newcommand{\fifrelingeant}{\textsc{Fifrelin4Geant4}\xspace}

\newcommand{\BaF}{\text{BaF$_2$}\xspace}
\newcommand{\faster}{\textsc{FASTER}\xspace}

\newcommand{\CRAB}{\textsc{Crab}\xspace}
\newcommand{\NUCLEUS}{\textsc{Nucleus}\xspace}
\newcommand{\CRESST}{\textsc{Cresst}\xspace}
\newcommand{\COSINUS}{\textsc{COSINUS}\xspace}
\newcommand{\TESSERACT}{\textsc{Tesseract}\xspace}
\newcommand{\RICOCHET}{\textsc{Ricochet}\xspace}
\newcommand{\EDELWEISS}{\textsc{Edelweiss}\xspace}

\newcommand{\CEvNS}{CE$\nu$NS\xspace}
\newcommand{\CaWO}{CaWO$_{4}$\xspace}
\newcommand{\AlO}{Al$_{2}$O$_{3}$\xspace}

\newcommand{\red}[1]{{\color{red}\textbf{[#1]}}}
\newcommand{\blue}[1]{{\color{blue}\textbf{[#1]}}}

\title{
The CRAB facility at the TU Wien TRIGA reactor: status and related physics program
}
        
\author{H.~Abele\thanksref{TUWien}
        \and P.~Ajello\thanksref{TUM}
        \and A.~Armatol\thanksref{IP2I}
        \and B.~Arnold\thanksref{HEPHY}
        \and J.~Billard\thanksref{IP2I}
        \and E.~Bossio\thanksref{IRFU}
        \and J.~Burkhart\thanksref{HEPHY}
        \and F.~Cappella\thanksref{INFN1}
        \and N.~Casali\thanksref{INFN1}
        \and R.~Cerulli\thanksref{INFN2, Roma1}
        \and J.~Colas\thanksref{IP2I}
        \and J-P.~Crocombette\thanksref{SRMP}
        \and G.~del~Castello\thanksref{Roma2,Roma3}
        \and M.~del~Gallo~Roccagiovine\thanksref{Roma2,Roma3}
        \and S.~Dorer\thanksref{TUWien,e1}
        \and C.~Doutre\thanksref{IRFU}
        \and A.~Erhart\thanksref{TUM,e1}
        \and S.~Fichtinger\thanksref{HEPHY}
        \and M.~Friedl\thanksref{HEPHY}
        \and P.~Garin\thanksref{IRFU}
        \and R.~Gergen\thanksref{TUWien}
        \and C.~Goupy\thanksref{IRFU}
        \and D.~Hainz\thanksref{TRIGA}
        \and D.~Hauff\thanksref{TUM,MPP}
        \and E.~Jericha\thanksref{TUWien}
        \and M.~Kaznacheeva\thanksref{TUM}
        \and H.~Kluck\thanksref{HEPHY}
        \and T.~Lasserre\thanksref{TUM,IRFU,MPIK}
        \and D.~Lhuillier\thanksref{IRFU} 
        \and O.~Litaize\thanksref{Cadarache} 
        \and P.~de~Marcillac\thanksref{IJCLab}
        \and S.~Marnieros\thanksref{IJCLab}
        \and R.~Martin\thanksref{IRFU,e1}
        \and E.~Namuth\thanksref{TUM}
        \and T.~Ortmann\thanksref{TUM}
        \and D.V.~Poda\thanksref{IJCLab}
        \and L.~Peters\thanksref{TUM,IRFU,MPIK}
        \and F.~Reindl\thanksref{TUWien,HEPHY}
        \and W.~Reindl\thanksref{HEPHY}
        \and F.~Rodari\thanksref{IRFU}
        \and J.~Rothe\thanksref{TUM}
        \and N.~Schermer\thanksref{TUM}
        \and J.~Schieck\thanksref{TUWien,HEPHY}
        \and S.~Sch\"{o}nert\thanksref{TUM}
        \and C.~Schwertner\thanksref{TUWien,HEPHY}
        \and G.~Soum-Sidikov\thanksref{IRFU}
        \and R.~Strauss\thanksref{TUM}
        \and R.~Thalmeier\thanksref{HEPHY}
        \and L.~Thulliez\thanksref{IRFU}
        \and C.~Trunner\thanksref{TRIGA}
        \and M.~Vignati\thanksref{Roma2,Roma3}
        \and M.~Vivier\thanksref{IRFU}
        \and P.~Wasser\thanksref{TUM}
        \and A.~Wex\thanksref{TUM}
}

\thankstext{e1}{emails: sebastian.dorer@tuwien.ac.at, andreas.erhart@tum.de, romain.martin2@cea.fr}



\institute{TU Wien, Atominstitut, 1020 Wien, Austria \label{TUWien}
           \and
           Physik-Department, Technische Universit\"{a}t M\"{u}nchen, D-85748 Garching, Germany \label{TUM}
           \and
           Univ. Lyon, Universit\'{e} Lyon 1, CNRS-IN2P3, IP2I-Lyon, Villeurbanne, F-69622, France \label{IP2I}
           \and
           Institut f\"ur Hochenergiephysik der \"Osterreichischen Akademie der Wissenschaften, A-1050 Wien, Austria \label{HEPHY}
           \and
           INFN, Sezione di Roma, I-00185, Roma, Italy \label{INFN1}
           \and
           INFN, Sezione di Roma "Tor Vergata", I-00133 Roma, Italy  \label{INFN2}
           \and
           Dipartimento di Fisica, Universit\`{a} di Roma "Tor Vergata", I-00133 Roma, Italy \label{Roma1}
           \and
           Universit\'e Paris-Saclay, CEA, DES, SRMP, F-91191 Gif-sur-Yvette, France \label{SRMP}
           \and
           Dipartimento di Fisica, Sapienza Universit\`{a} di Roma, I-00185 Roma, Italy \label{Roma2}
           \and
           Istituto Nazionale di Fisica Nucleare, Sezione di Roma, I-00185 Roma, Italy \label{Roma3}
           \and
           IRFU, CEA, Universit\'e Paris-Saclay, 91191 Gif-sur-Yvette, France \label{IRFU}
           \and
           TU Wien, TRIGA Center Atominstitut, 1020 Wien, Austria \label{TRIGA}
           \and
           Max-Planck-Insitut f\"{u}r Physik, D-80805 M\"{u}nchen, Germany \label{MPP}
           \and
           CEA, DES, IRESNE, DER, Cadarache, F-13108 Saint-Paul-Lez-Durance, France \label{Cadarache}
           \and
           Universit\'{e} Paris-Saclay, CNRS/IN2P3, IJCLab, 91405 Orsay, France \label{IJCLab}
           \and
           Max-Planck-Institut f\"ur Kernphysik, Saupfercheckweg 1, Heidelberg, 69117, Germany \label{MPIK}
}

\date{Received: date / Accepted: date}

\maketitle


\begin{abstract}

The \CRAB (Calibrated nuclear Recoils for Accurate Bolometry) project aims to precisely characterize the response of cryogenic detectors to sub-keV nuclear recoils of direct interest for coherent neutrino-nucleus scattering and dark matter search experiments. The \CRAB method relies on the radiative capture of thermal neutrons in the target detector, resulting in a nuclear recoil with a well-defined energy. We present a new experimental setup installed at the TRIGA Mark-II reactor at Atominstitut (Vienna), providing a low intensity beam of thermal neutrons sent to the target cryogenic detector mounted inside a wet dilution refrigerator Kelvinox~100. A crown of \BaF detectors installed outside the dewar enables coincident detection of the high-energy $\gamma$ escaping the target crystal after neutron capture. After the presentation of all components of the setup we report the analysis of first commissioning data with a \CaWO detector of the \NUCLEUS experiment. They show stable operation of the cryostat and detectors on a week-scale. Due to an energy resolution currently limited to 20~eV we use neutron beam induced events at high energy, in the 10 to 100~keV range, to demonstrate the excellent agreement between the data and simulation and the accurate understanding of external background. Thanks to these data we also propose an updated decay scheme of the low-lying excited states of $^{187}$W. Finally, we present the first evidence of neutron-capture induced coincidences between $\gamma$-detectors and a cryogenic detector. These promising results pave the way for an extensive physics program with various detector materials, like \CaWO, \AlO, Ge and Si.

\end{abstract}

\section{Introduction}
\label{sec:intro}

Recent advances in the detection of very low-energy nuclear recoils, down to the sub-keV energy scale, have opened up new perspectives in the study of coherent elastic neutrino-nucleus scattering (\CEvNS) and in the search for light dark matter (DM). However, the interactions of the primary knock-on atom (PKA) with the crystal lattice of cryogenic detectors are complex. The distribution of energy losses between the various channels --- ionization, atomic collisions, scintillation, heat and creation of crystal defects --- is highly dependent on the initial energy of the PKA and the detector material. Calibrations with photon sources induce electronic recoils, for which the distribution of energy losses in the various channels mentioned above will be very different from that of nuclear recoils, at work in neutrino or DM scattering processes. Also, depending on the calibration method, the recoiling particles are not always uniformly distributed throughout the detector volume. All these effects result in significant biases and/or uncertainties in the energy reconstruction of \CEvNS and DM signals. 

In this context, the \CRAB (Calibrated nuclear Recoils for Accurate Bolometry) collaboration offers high-precision measurements of nuclear recoils of known energy, distributed throughout the detector volume and in the energy range of interest for neutrino and DM physics. The principle of this calibration method is based on the capture of thermal neutrons (kinetic energy of about 0.025~eV) by nuclei of a cryogenic detector \cite{Thulliez_2021}. The compound nucleus formed decays with a $\gamma$-cascade of total energy $S_\text{n}$, the neutron separation energy. The $S_\text{n}$ value is specific to the target nucleus and typically between 5 and 11~MeV. When the cascade consists of a single $\gamma$-ray, the induced recoil energy $E_\text{r}$ of the emitting nucleus of mass $M_\text{n}$ is simply $E_\text{r} = S_\text{n}^2 / 2M_\text{n}$. If the nuclear properties are favorable (large natural abundance of the target nuclide, large neutron capture cross-section and large probability of single-$\gamma$ de-excitation), then a calibration peak will be measurable in the induced recoil spectrum. With typical dimensions of a few mm to a few cm, neutron captures are evenly distributed throughout the volume of a cryogenic detector. This small size also allows the high-energy $\gamma$-ray to escape easily from the detector without any energy deposition, preserving the signal of the calibrated nuclear recoil. This method has been validated experimentally for the first time with a \CaWO cryogenic detector from the \NUCLEUS experiment \cite{NUCLEUS:2019igx}. This material is particularly favourable for the \CRAB method, thanks to the nuclear properties of tungsten, which is also the main target for the \CEvNS events that \NUCLEUS wants to study. The main calibration peak, expected at 112~eV, was observed by exposing the cryostat to a neutron flux from a commercial $^{252}$Cf fission source in polyethylene placed nearby \cite{PhysRevLett.130.211802}. This result was confirmed by a similar measurement using \CaWO detectors of the \CRESST experiment, which searches for DM events at the Laboratori Nazionali del Gran Sasso in Italy \cite{PhysRevD.108.022005}.

However, these measurements suffer from significant background due to the fact that the neutron flux reaching the cryogenic detector is low, spread over the whole cryostat and contaminated by fast neutrons and $\gamma$-rays from the source. In this paper we present a new experimental configuration of the \CRAB project, optimized for high-precision measurements of the mean position of calibration peaks with high signal-to-background ratio. The neutron source is now a pure and collimated beam of thermal neutrons provided by a nuclear research reactor and $\gamma$-detectors around the cryostat allow the detection in coincidence of the nuclear recoil and the emitted $\gamma$-ray(s).  Section~\ref{sec:overview} presents a brief overview of the setup recently installed on the reactor site. We then describe in more detail the various components of the experiment: the neutron beam in Sect.~\ref{sec:beam}, the cryostat infrastructure in Sect.~\ref{sec:cryostat}, the cryogenic detectors in Sect.~\ref{sec:cryo_det} and the $\gamma$-detectors in Sect.~\ref{sec:gamma_det}. In Sect.~\ref{subsec:simulation} we present the simulation package used to predict all the events induced by the neutron beam in the cryogenic detectors. We show in Sect.~\ref{sec:commissioning} that these predictions, supplemented by independent measurements of the external background, are found to be in very good agreement with the data collected during the commissioning of the experiment. We also present evidence of the coincident detection of a neutron capture-induced event in a cryogenic detector and a high-energy $\gamma$-ray from the same de-excitation cascade in a $\gamma$-detector outside the cryostat. To our knowledge, the proof of concept of this technique is a world premiere, opening up exciting new measurement prospects.
Sect.~\ref{sec:physics} concludes this article with a discussion of the physics program envisaged with this setup. We show that \CRAB measurements are applicable to most of the detector materials used by the community, with original applications in solid state and nuclear physics. 

\section{Overview of the experimental site}
\label{sec:overview}

\begin{figure*}[t!]
    \centering
    \includegraphics[width=\linewidth]{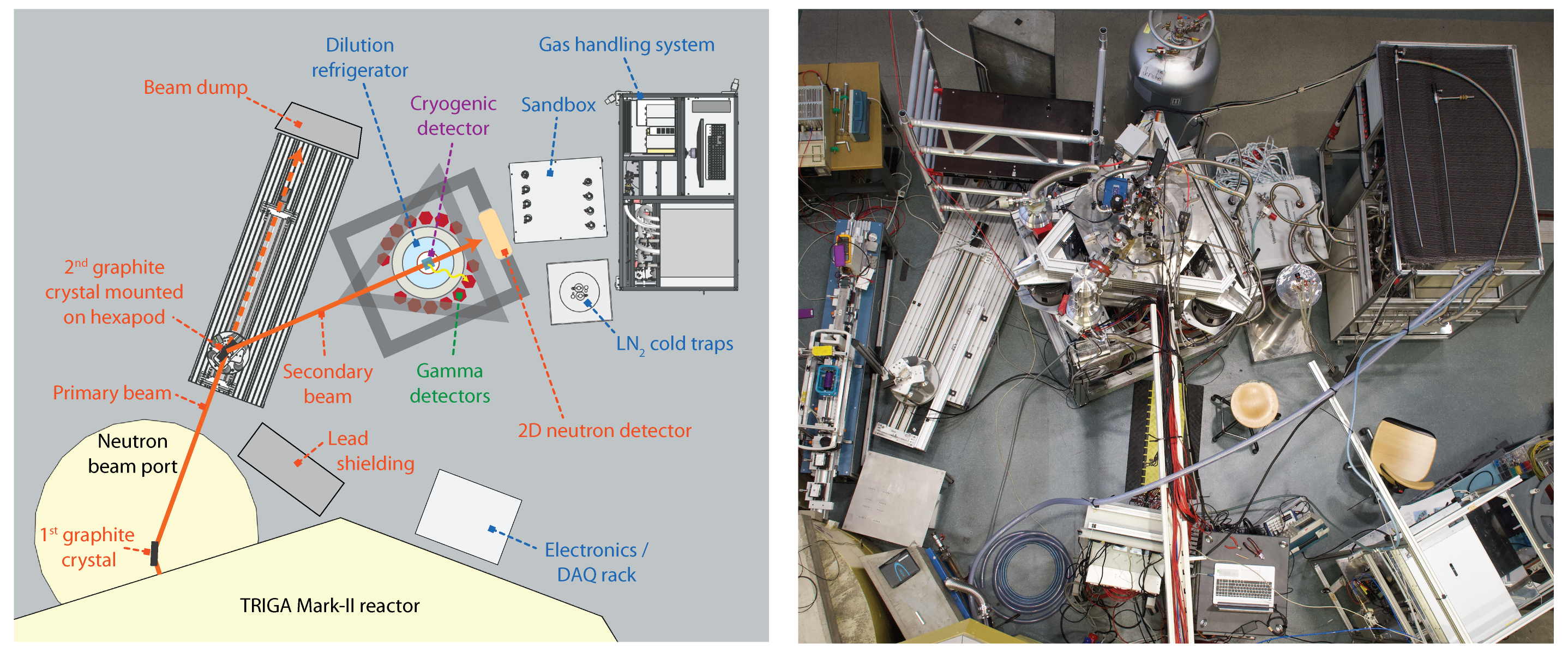}
    \caption{Schematic and photography of the \CRAB facility at the TRIGA Mark-II reactor in Vienna. The left panel shows a schematic top view of the experiment with the neutron beam line (orange labels), the cryogenic infrastructure (blue labels), the cryogenic detector (purple labels) and the $\gamma$-detectors (green labels). The right panel shows an image of the setup taken from the balcony of the reactor.}
    \label{fig:Overview}
\end{figure*}

The various nuclear processes at work in neutron sources (fission, spallation, alpha capture, etc.) produce neutrons in the MeV energy range, well above the thermal energies requested for \CRAB and are often accompanied by $\gamma$-rays. Obtaining a pure thermal neutron flux requires to slow down and thermalize the produced neutrons to 25~meV in a moderator and to block the gamma radiations with an extensive shielding. These conditions are typically available from research reactors, from which one can extract very pure and collimated thermal neutron beams. For the \CRAB measurements, we chose the TRIGA-Mark II reactor at the Atominstitut (ATI) in Vienna, which offered an available site with a dedicated neutron line. The low power of this reactor, 250~kW$_{th}$, is also well suited to data taking with cryogenic detectors, whose slow time response typically limits their total counting rate to around 1~Hz. With detector sizes in the cm range, the required neutron fluxes are in the 10 to 1000~cm$^{-2}$.s$^{-1}$ range, compatible with ATI neutron beams. Figure~\ref{fig:Overview} shows an overview of the experimental setup. The primary beam from the reactor is reflected on a graphite crystal toward a dilution fridge, mounted on a supporting structure. The gas handling system (GHS) is organized in a compact and transportable rack, designed for measurements on the reactor site. The neutron beam is sent directly through the dewar containing the helium bath and through the walls of the cryostat insert. 

The system is completed by a ring of $\gamma$-detectors installed around the cryostat. With several MeV energy, the $\gamma$-rays responsible for the calibrated recoils not only escape easily from the cryogenic detector, but can also go through the thin walls of the cryostat. The detection of the $\gamma$-ray in coincidence with the nuclear recoil in the cryogenic detector has great physics potential, enabling to tag the recoil in time, energy and direction (see Sect. \ref{sec:physics}).

\section{Neutron beam from the Vienna research reactor}
\label{sec:beam}

\subsection{TRIGA Mark-II reactor}
\label{subsec:reactor}

The TRIGA Mark-II nuclear research reactor at ATI operates at a nominal power of 250~kW$_{\text{th}}$ for seven hours per day at approximately 220 days per year. 
Its core has a maximum capacity of 85 uranium-zirconium-hydride composite fuel elements and generates a thermal neutron flux of approximately $10^{13}$~cm$^{-2}$s$^{-1}$ at its center. 
The reactor supplies neutrons to multiple experiments via four beam tubes, a thermal column, a dry irradiation room, a pneumatic conveying system, and irradiation tubes for samples in and around the reactor core. 
Being the only beam tube that is not facing the core directly, but tangentially, beam tube D provides a lower $\gamma$ and fast neutron background than the other tubes and is home to the experimental setup of \CRAB \cite{ati-triga-center-td, ati-triga-center-rk, ati-triga-center-beschreibung}.

\subsection{\CRAB neutron line}
\label{subsec:crab_line}

\begin{figure}[htbp]
    \centering
    \includegraphics[width=0.8\linewidth]{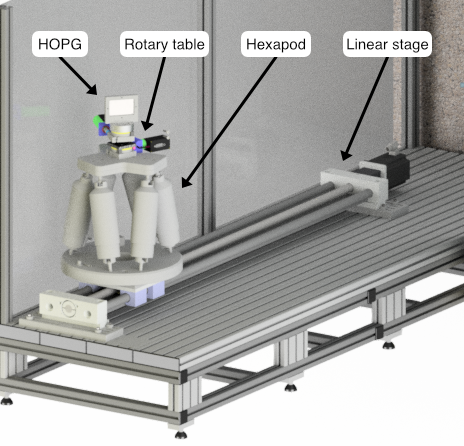}
    \caption{CAD drawing of the mechanical components used for the precise positioning of the selector graphite crystal.}
    \label{fig:neutrons-mechanical-setup}
\end{figure}

In order to transport thermal neutrons from the reactor to the cryogenic detector, the neutron beam is reflected on two graphite crystals (see Fig.~\ref{fig:Overview}). 
The first crystal, the monochromator, is mounted in a fixed position behind a concrete radiation shielding structure. 
The second crystal, the selector, is located outside the shielding and is mounted on an M-037.DG rotary table from Physik Instrumente \cite{PI-drehtisch-ws}. 
The rotary table sits on top of an H820 6-axis hexapod from Physik Instrumente \cite{PI-hexapod-ws}, that itself is mounted on a DL-SLW linear stage from igus \cite{igus-linear-stage}. 
This setup allows for the precise positioning of the selector crystal within the neutron beam in order to align the beam with the cryogenic detector target. An overview of the setup is depicted in Fig.~\ref{fig:neutrons-mechanical-setup}. 
Furthermore, a DENEX 200-TN 2-dimensional position sensitive $^3$He neutron detector \cite{DENEX-2D-PSD-ws} is mounted behind the cryostat and is used as a qualitative imaging tool to check the alignment of the neutron beam with the cryogenic detector (see Sect. \ref{subsec:align}). Vertices of neutron captures on $^3$He atoms in this multi wire proportional chamber can be reconstructed with a FWHM spatial resolution on the millimeter scale in both horizontal and vertical directions due to two perpendicular grids of gold coated tungsten-rhenium-alloy wires. 

Quantitative neutron flux measurements are performed with a VacuTec 70 063 \cite{vacutec-he3-counter} proportional $^3$He counter tube. 
The setup also allows the usage of different boron-loaded apertures to further tailor the neutron beam if necessary. 
 
The reactor's beam tube D provides a thermal neutron beam with an energy following a Maxwell-Boltzmann distribution for $T \approx 300$~K. 
In order to supply more than one experiment with neutrons from the same beam tube, three highly oriented pyrolytic graphite (HOPG) monochromator crystals are positioned in succession in the initial beam. 
These crystals are rotated at different angles $\theta$ along the vertical axis and therefore reflect different neutron wavelengths $\lambda$ of the beam given by Bragg's reflection law \cite{bragg:reflection-by-crystals}
\begin{equation}
    n \lambda = 2 d \sin{\left(\theta\right)}
\end{equation}
with an angle of $2 \theta$ out of the initial beam, $d$ the crystal lattice spacing and the diffraction order $n \in \mathbb{N}$.
The \CRAB experiment's beam line is supplied by a monochromator crystal at 23.17$^\circ$ (1$^{\text{st}}$ graphite crystal in Fig.~\ref{fig:Overview}) and thus contains neutrons with \mbox{$\lambda_n \approx 2.64 / n$ \AA}. 
Its composition mainly consists of neutrons with wavelengths of 0.88 \AA, 1.32 \AA \ and 2.64 \AA. 
In the following section, this beam will be called the primary beam.

\begin{figure}[htbp]
    \centering
    \includegraphics[width=0.9\linewidth]{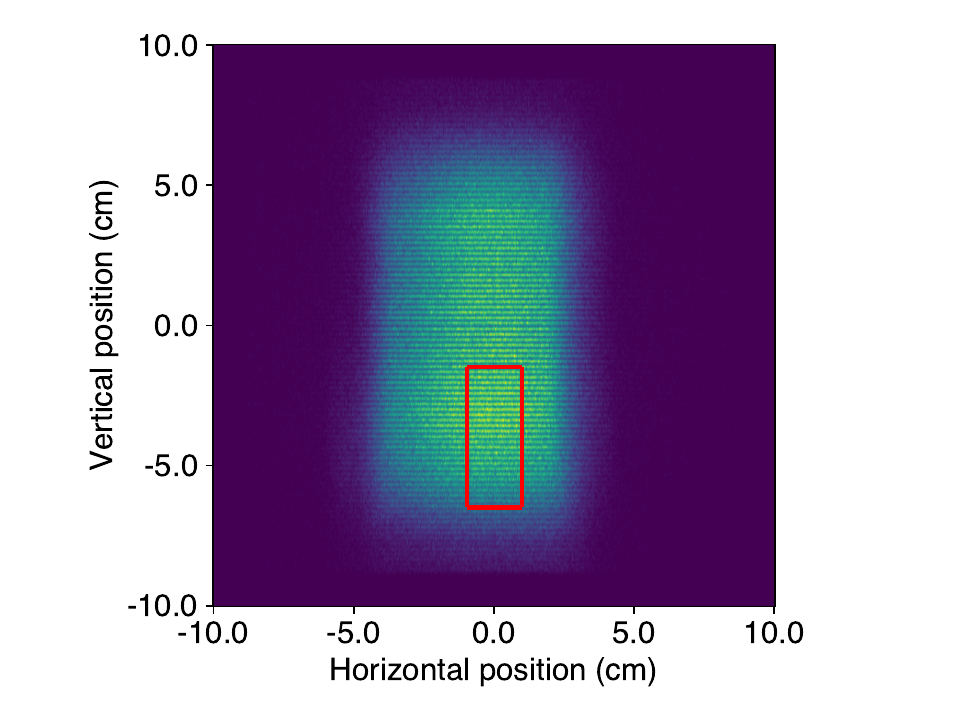}
    \caption{Spatial distribution of the primary neutron beam at the same distance from the neutron guide as the selector graphite crystal in the \CRAB configuration. The red box outlines the position of the selector crystal in the beam. The horizontal lines originate from shadows casted by one wire layer of the neutron detector's multi-wire proportional chamber design.}
    \label{fig:primary-beam-2d-at-crystal}
\end{figure}

Figure~\ref{fig:primary-beam-2d-at-crystal} shows a 2-dimensional image of the primary neutron beam at the same distance and position at which the selector crystal is placed (109 cm away from the beam guide). 
The size of the beam is defined via the full width at half maximum (FWHM) of the horizontal and vertical projections of the 2-dimensional data and is around $6.3\times12.0$~cm$^{2}$ (width~$\times$~height). 
Furthermore, the beam displays an inhomogeneous shape both horizontally and vertically. 
In the horizontal direction, the beam has a negative skew with respect to the neutron flight direction. 
Vertically, the maximum neutron flux is located near the bottom of the beam, with an approximately 15\% lower plateau in the middle and upper region. 
The measured FWHM divergence is $^{+1.28^\circ}_{-1.45^\circ}$ and $^{+0.68^\circ}_{-0.90^\circ}$ for the vertical and horizontal directions, respectively. 
The neutron flux averaged over an area of $2\times4.5$~cm$^{2}$ at the position of the selector crystal was measured with a $^3$He neutron counter tube and is $\phi \approx (4296 \pm 430)$ cm$^{-2}$s$^{-1}$. 

As described above, the selector crystal is placed in the primary beam at a rotation angle of 23.17$^\circ$ in order to reflect the neutron beam precisely onto the cryogenic detector target inside the cryostat.
Three main points were taken into consideration for the choice of the rotation angle of this crystal. 
First, the neutron flux in the secondary beam should be high enough to provide margins for the implementation of neutron optical components for beam shaping and for the collection of calibration data within reasonable time. 
Second, the cryostat should be positioned away from the primary beam in order to reduce the neutron and $\gamma$ background. 
Third, the spatial constraints imposed by the experimental area. 
Without additional apertures, this secondary neutron beam in front of the cryostat has dimensions of around $3.0\times6.4$~cm$^{2}$. 
The divergence is calculated to be $^{+0.32^\circ}_{-0.34^\circ}$ for the vertical direction and $^{+0.24^\circ}_{-0.69^\circ}$ for the horizontal one. 
The neutron flux averaged over an area of $2\times~4.5$~cm$^{2}$ in front of the cryostat is measured to be $\phi~\approx~(469~\pm~47)$~cm$^{-2}$s$^{-1}$. 
The neutron transmission through the whole cryostat (including the insert, cryogenic detectors and two layers of $\mu$-metal at the exit surface of the dewar, but without liquid helium) is measured to be around (6.71 $\pm$ 0.67)\% of the incident flux. 
By approximating the transmission factor for the liquid helium volume with 0.64 and with 0.89 for the $\mu$-metal sheets, the flux at the position of the cryogenic detectors inside the cryostat can be estimated to be $\phi~\approx~(121~\pm$~17)~cm$^{-2}$s$^{-1}$.
Due to an inhomogeneous beam shape, there are significant deviations in the local flux from the average given above.

\section{Cryogenic infrastructure}
\label{sec:cryostat}

\begin{figure}[ht!]
    \centering
    \includegraphics[width=.9\linewidth]{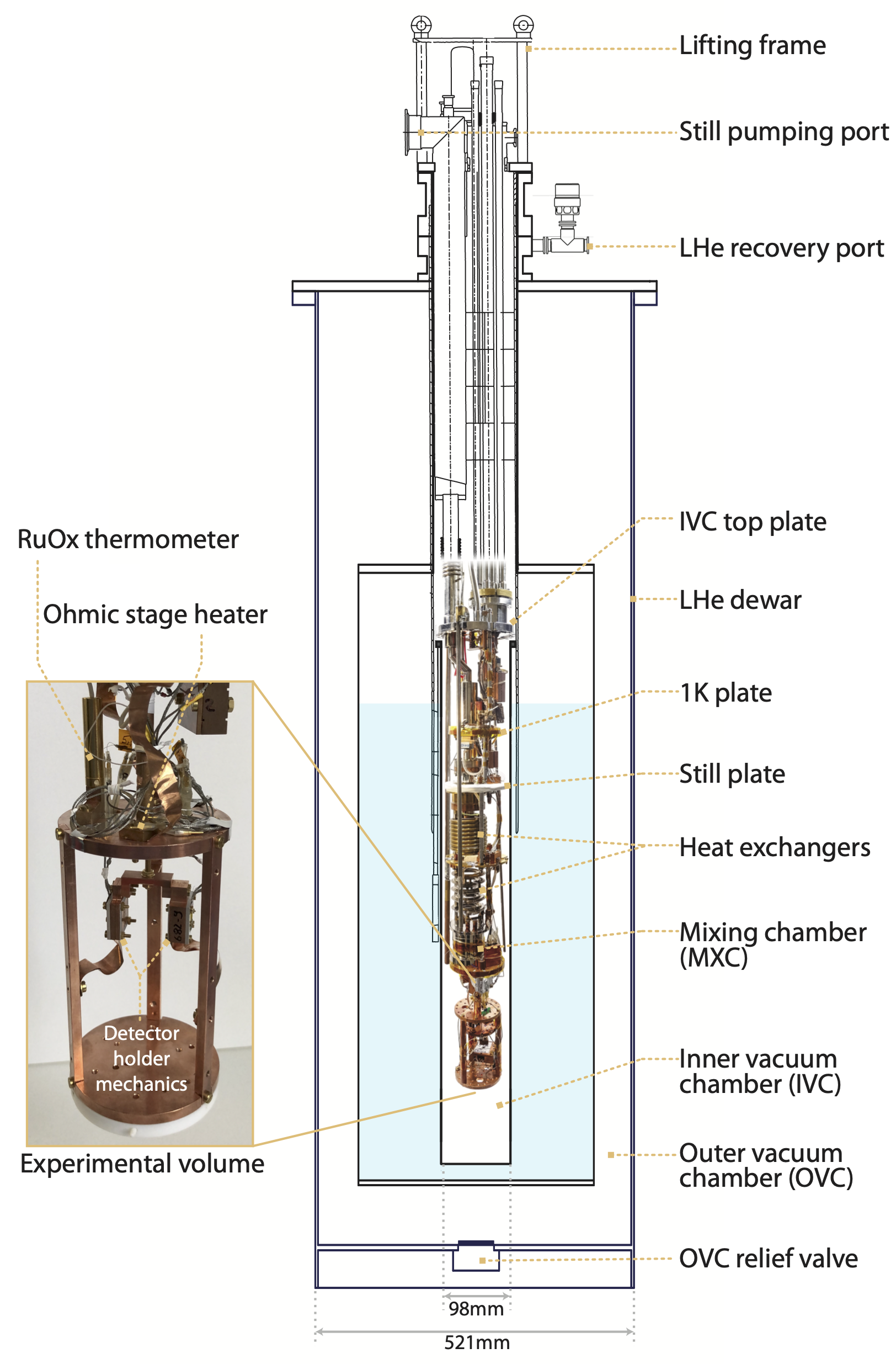} %
     \caption{Schematic overview of the wet $^3$He-$^4$He dilution refrigerator Kelvinox 100 from Oxford Instruments plc. The cryostat insert is enclosed by the inner vacuum chamber (IVC) and inserted in a dewar containing up to 60~l of LHe. The cylindrical experimental volume (130~mm height, 80~mm diameter) is aligned with the neutron beam.}
    \label{fig:cryostat_scheme}
\end{figure}

The cryostat is a wet $^3$He-$^4$He dilution refrigerator Kelvinox~100 from Oxford Instruments plc consisting of an insert in the center of a dewar that can contain up to 60~l of liquid helium (see Fig.~\ref{fig:cryostat_scheme}). Apart from weekly, few-hour periods of liquid helium refilling, long-term measurements with cryogenic detectors are possible over a period of several months under very stable conditions.

\subsection{Support structure}
\label{subsec:support}

\begin{figure}[ht!]
    \centering
    \includegraphics[width=\linewidth]{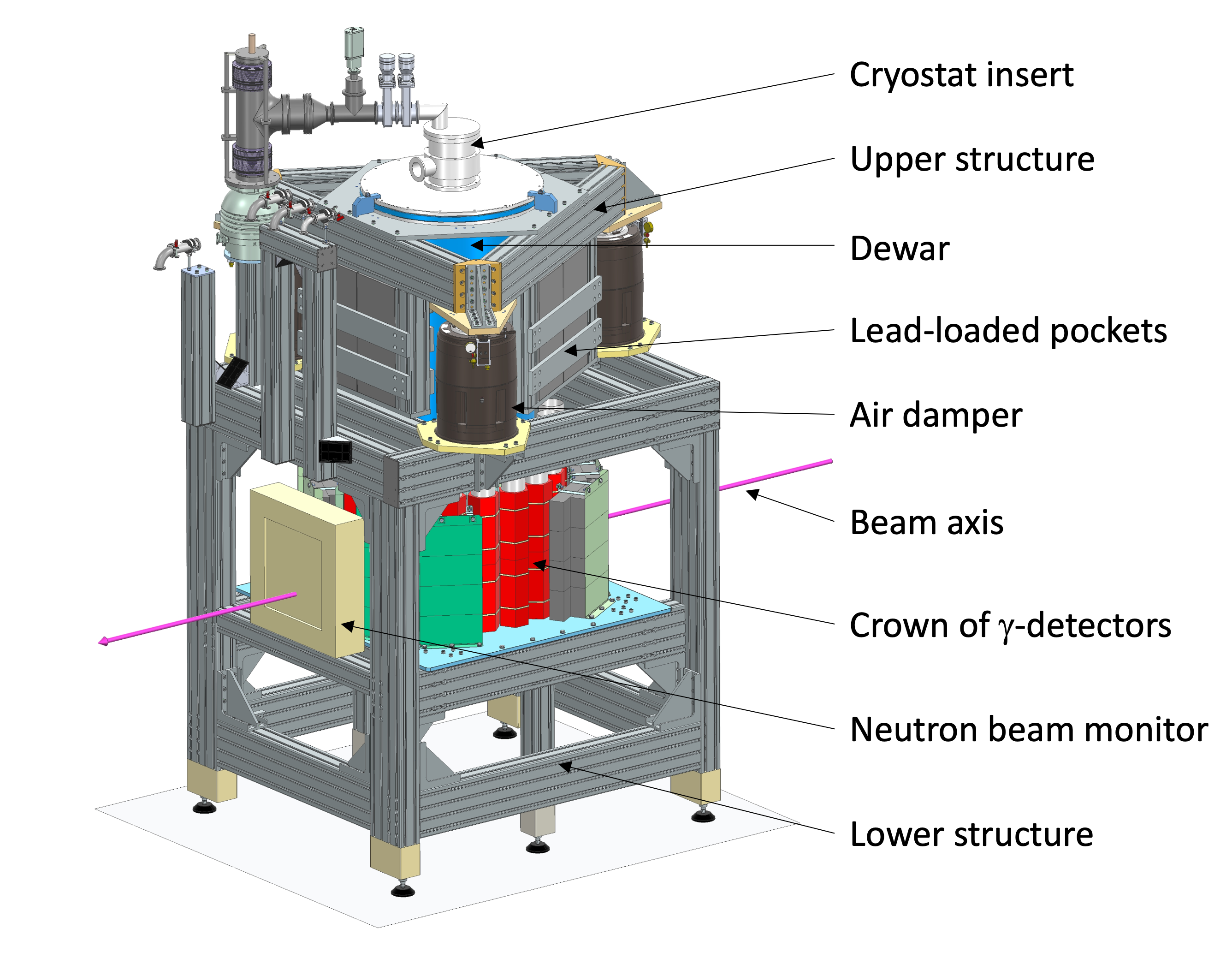}
    \caption{Illustration of the support structure holding the cryostat, the $\gamma$-detectors and the neutron beam monitor. Part of the lead shielding around the $\gamma$-detector is cut out for better visibility.}
    \label{fig:Support_Structure}
\end{figure}

The cryostat insert and its dewar are suspended from a triangular structure of aluminum profiles by Alv\'aris \cite{Alvaris}. This upper structure rests on a square-base lower structure \textit{via} Newport S-2000 vibration isolators \cite{Newport_S2000}, located under each corner of the triangle (Fig. \ref{fig:Support_Structure}). For optimum operation of these air dampers, the suspended mass, initially 250~kg, is increased to around 500 kg by means of compartments fixed under each side of the upper triangle and filled with lead bricks. Isolation of the cryogenic system from external vibrations is completed by a nearby sand box, through which all the flexible hoses connecting the cryostat to the pumps in the GHS pass. The cryostat is also electrically isolated from this metal structure by a plastic ring at the top flange of the dewar to avoid ground loops between the different components of the experiment. Halfway up to the lower structure, a plate supports the $\gamma$-detectors and their shielding, arranged in a ring around the dewar (see Sect. \ref{sec:gamma_det}) and representing a mass of around 1~tonne. The 2D-neutron detector described in Sect. \ref{sec:beam} is installed downstream of the dewar at the same height, centered on the beam axis. 

\subsection{Kelvinox 100 dilution refrigerator}
\label{subsec:wet_cryo}

\subsubsection{Characterization of the material and thickness of the dewar walls}

\begin{figure}[ht!]
    \centering
    \includegraphics[width=1\linewidth]{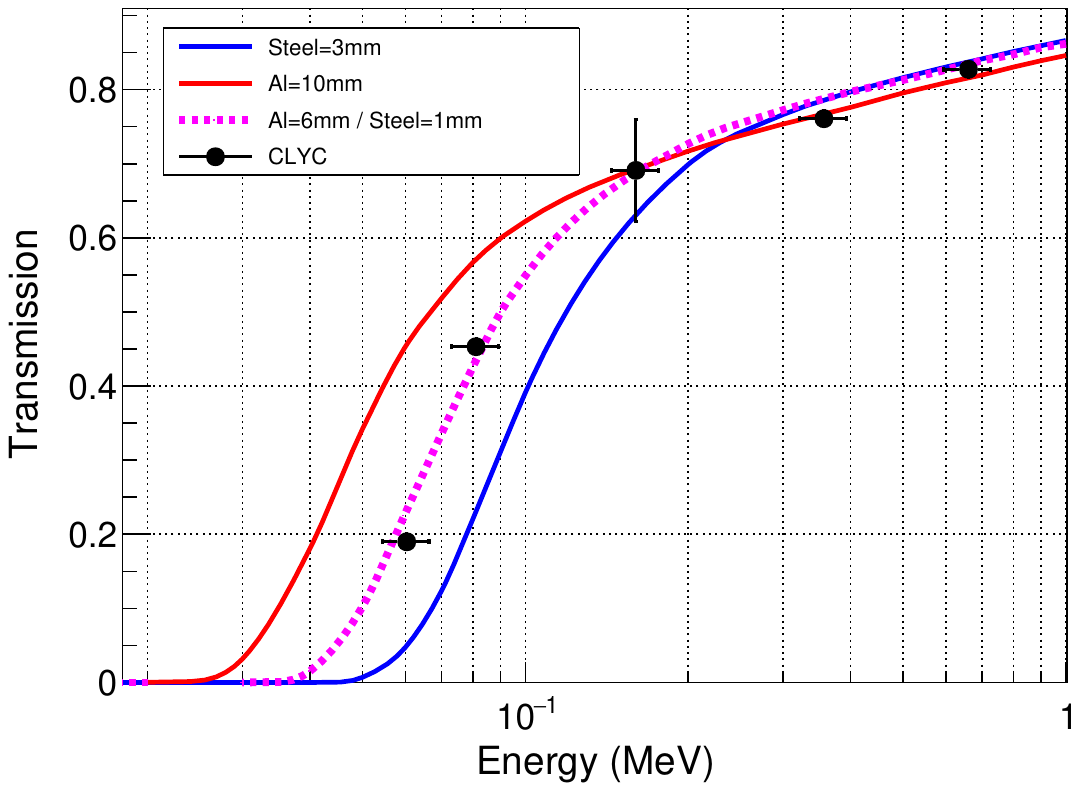}
    \caption{Attenuation factors of $\gamma$-rays as measured by the CLYC detector (black points) are compared to \geant simulations with different materials and thicknesses of the dewar walls. Only statistical uncertainties are displayed. They are computed from the uncertainties of the fitted mean position of the detected gamma lines and are found to be very small except for the low intensity peak at 160.6 keV from $^{\text{133}}$Ba. Only a mix of aluminum and steel walls can match the data.
    }
    \label{fig:GammaTransmission}
\end{figure}

The Kelvinox~100 cryostat dates back from the 1990's and precise information on the thickness and material composition of the dewar internal layers is uncertain while it is crucial for our high precision measurements since the thermal neutron beam is sent directly through the cryostat to reach the target detector, with no dedicated dewar entrance or exit window. Therefore we performed $\gamma$-attenuation measurements to supplement the scarce information from the manufacturer.\\
\indent Different radioactive sources were inserted into the empty dewar at the position of the detector module thanks to a long U-shaped structure made of aluminum profiles. The same structure supported a CLYC detector (Cs$_2$LiYCl$_6$:Ce scintillator \cite{CLYCref}) outside the dewar, at a fixed distance from the sources. 
Attenuation factors were then obtained by comparing count rates with and without the dewar in place. Five $\gamma$-lines were measured from the following isotopes: $^{\text{137}}$Cs (661.7~keV), $^{\text{133}}$Ba (356.0~keV, 160.6~keV, 81.0~keV) and $^\text{241}$Am (59.5~keV).
In Fig.~\ref{fig:GammaTransmission} the corresponding attenuation factors are compared to attenuation curves calculated for various combinations of materials (5083 aluminium and 316 stainless steel) and thicknesses using \geant with the Livermore physics list. 
The shape of the experimental curve turns out to be discriminant, clearly excluding the presence of a single material in the walls of the dewar. A combination of a 6~mm 5083 aluminum layer and a 1~mm 316 stainless steel layer best describes the data, in good agreement with manufacturer's estimate.

The last component of the dewar to be characterized is the thermal insulation, made of many interleaved layers of mylar and aluminized mylar surrounding the LHe vessel. This low density material has a negligible impact on the $\gamma$-transmission measurements but it affects the neutron beam due to its hydrogen-rich composition. An intermediate layer of mylar, defined as Polyester\_C10H8O4 in the thermal neutron scattering library \ncrystal \cite{caiKittelmann2020_ncrystal, KittelmannCai2021_ncrystal}, is therefore added to the geometry of the \geant simulation of the experiment (see Sect. \ref{subsec:simulation}). A total thickness of 2~mm gives the best agreement with the total attenuation factor of the neutron beam reported in Sect. \ref{sec:beam}. This thickness is in line with the manufacturer's data of around 150 sheets of mylar+aluminized mylar being between 12 and 25~$\mu$m thick. In conclusion, the dewar is made of 6~mm of aluminum, 2~mm of mylar and 1~mm of 316 stainless steel. Additional cryostat-internal shielding layers include the vessel of the IVC, composed of 2~mm stainless steel and 1~mm copper, as well as the 50\,mK radiation shield, which consists of 1~mm copper.

\subsubsection{Schematic overview of the wet cryostat}

The cryostat has a cooling power of $\sim$\,140~$\mu$W at 100~mK and reaches a base temperature below~10~mK. The still and mixing chamber (MXC) plates of the cryostat are equipped with ohmic heaters and can be temperature-stabilized by the use of a Proportional Integral Derivative (PID) loop. Two calibrated diodes with a wide dynamic temperature range from 300~K down to 1~K (read by a Lakeshore 325 temperature controller) are used to measure the temperature on the 4~K stage and of the 1~K pot during cool-down. Two ruthenium oxide (RuOx) thermometers from Bluefors read by a Lakeshore 372 AC resistance bridge are used to monitor the temperature of the MXC stage and the detector box during continuous operation. These RuOx thermometers are calibrated down to 7~mK.

The cylindrical experimental volume below the mixing chamber has a diameter of 130~mm and a height of 80~mm. Several detectors can be mounted on two NOSV copper plates. The volume is enclosed by a 0.3~mm copper MXC shield to block thermal radiation from warmer stages.

The cryostat is optimized for the operation of Transition Edge Sensor (TES) based cryogenic detectors and is equipped with a 3-channel DC-SQUID (Superconducting Quantum Interference Device) system from Magnicon. The custom-built TES-readout circuit is implemented with bias and heater lines filtered at 4~K and shunt resistors mounted on the MXC stage. The cabling is realized using twisted pair Nb/Ti cables thermalized at multiple temperature stages. The cryostat is also equipped with optical fibers to perform energy calibrations of cryogenic detectors using a burst of light pulses emitted by an LED operated at room temperature.

\subsubsection{Design and implementation of a compact GHS}

A compact and mobile gas handling rack has been designed for use at the reactor site to manage the circulation of helium gas in the system. The GHS comprises the central KelvinoxIGH -- a commercial intelligent gas handling unit from Oxford Instruments plc -- with internal valves and pressure gauges to control the flow of $^3$He and $^4$He, which can be viewed and operated remotely \textit{via} control software. The GHS also includes a 1 K pot pumping system driven by two multi-stage roots pumps (Pfeiffer ACP\,40) operating in series. The $^3$He-$^4$He mixture is circulated by a turbomolecular drag pump (Pfeiffer HiPace\,400) mounted close to the still line outlet on the cryostat insert rack to maximize pumping power and supported by a multi-stage roots pump (Pfeiffer ACP\,40) mounted in the GHS system. An additional turbomolecular pumping station, consisting of a Pfeiffer HiPace 400 turbomolecular pump and a Pfeiffer ACP\,40 backup pump, is used to evacuate the inner vacuum chamber (IVC) prior to cool-down. The system is equipped with eight additional pressure gauges (Pfeiffer PCR\,280) for accurate pressure monitoring and a LHe level meter to monitor the LHe level in the LHe dewar.

\subsection{Beam alignment}
\label{subsec:align}
\begin{figure}[ht!]
    \centering
    \includegraphics[width=\linewidth]{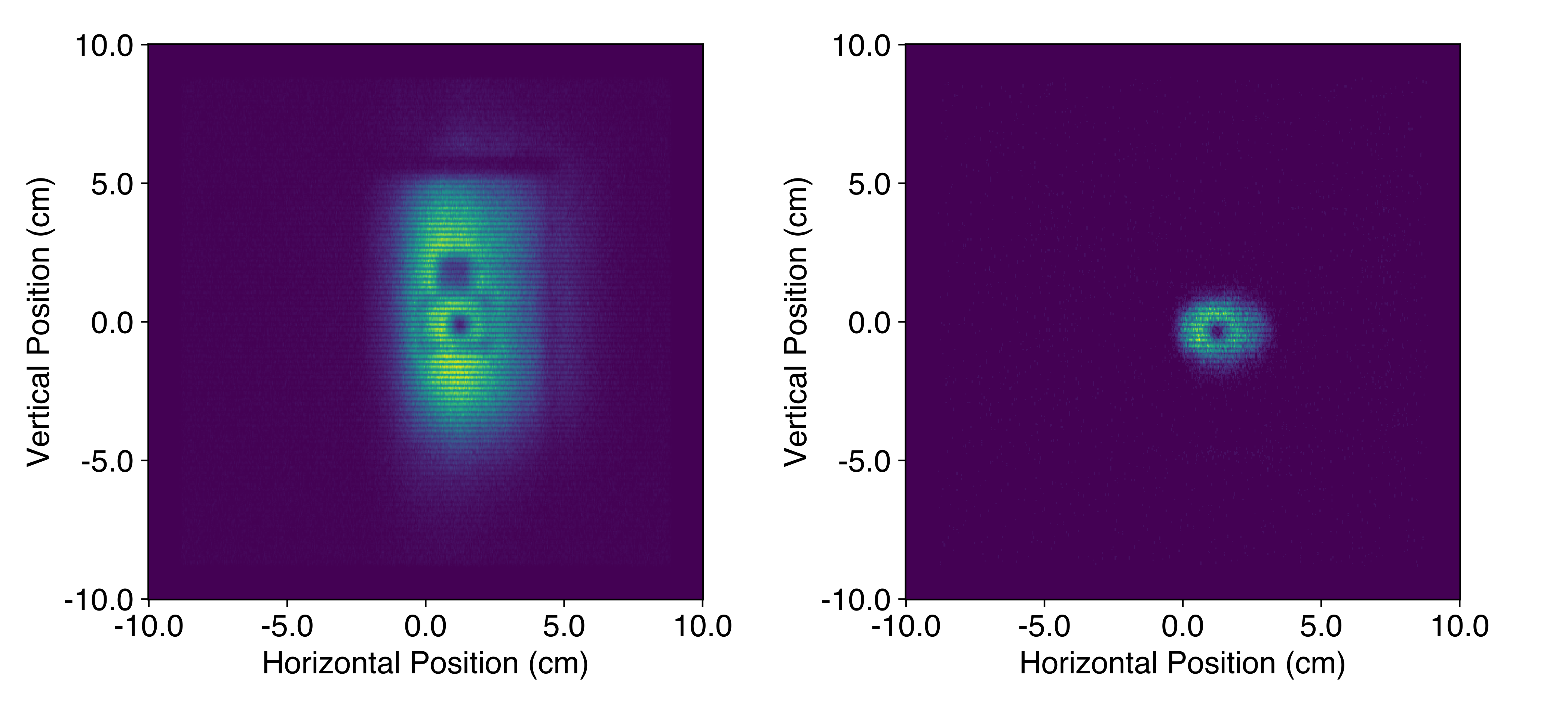}
    \caption{Left: image of the beam spot transmitted through the cryostat. The shadow of the copper holder described in Sect. \ref{sec:cryo_det} as well as the central square footprint of the $^6$LiF crystal are clearly visible. Right: image of the same experimental configuration with a 2~cm wide circular aperture in place. Most of the walls of the copper holder are masked and the aperture is well centered on the crystal.}
    \label{fig:Beam_Spots}
\end{figure}

Alignment of the entire cryostat assembly with the neutron beam axis is carried out in two stages. First, the cryostat insert is installed without the dewar, using an adapter flange reproducing the detector box position within millimeter precision. This allows a direct view, at room temperature, of a dummy cryogenic detector consisting of a $^6$LiF crystal. The $^6$LiF is 95\% enriched in $^6$Li, hence totally opaque to thermal neutrons. Thanks to its dimensions of $5.28\times6.27\times3.56$~mm$^{3}$, comparable to those of the \CaWO crystals to be calibrated by the next \CRAB measurements, the same cryogenic detector supports can be used, with the smallest dimension oriented along the beam axis. In reactor-ON mode, it casts a clear shadow pattern on the 2D-neutron monitor within the neutron beam spot allowing an online final adjustment of the beam axis with the hexapod. In a second stage the dewar is placed back around the insert and it is checked that the projected shadow of the $^6$LiF crystal is still observed at the same position, with millimeter precision. Reference images of the transmitted beam spot are taken to validate the size and position of a collimator installed upstream of the cryostat to reduce the beam spot extension to an area comparable to that of the target crystal (see Fig.~\ref{fig:Beam_Spots}). This collimator consists of a 5~mm thick rubber plate loaded with B$_4$C, into which a 2~cm wide circular aperture has been cut. The size of this aperture, reconstructed 80~cm downstream in the neutron monitor is $3.5\times2.9$~cm$^{2}$ (w $\times$ h), confirming the small divergence of the neutron beam. 
Comparing images obtained with the cryostat at room temperature and cooled down without any collimator, we see no evidence of a shift in detector position due to thermal shrinkage, and set an upper limit of 0.4~mm, the size of a pixel, on the variation of the detector position.

\section{Cryogenic detectors}
\label{sec:cryo_det}

\subsection{Benchmark performance measurements}
\label{sec:det_benchmark}
\begin{figure}[ht!]
    \centering
    \includegraphics[width=\linewidth]{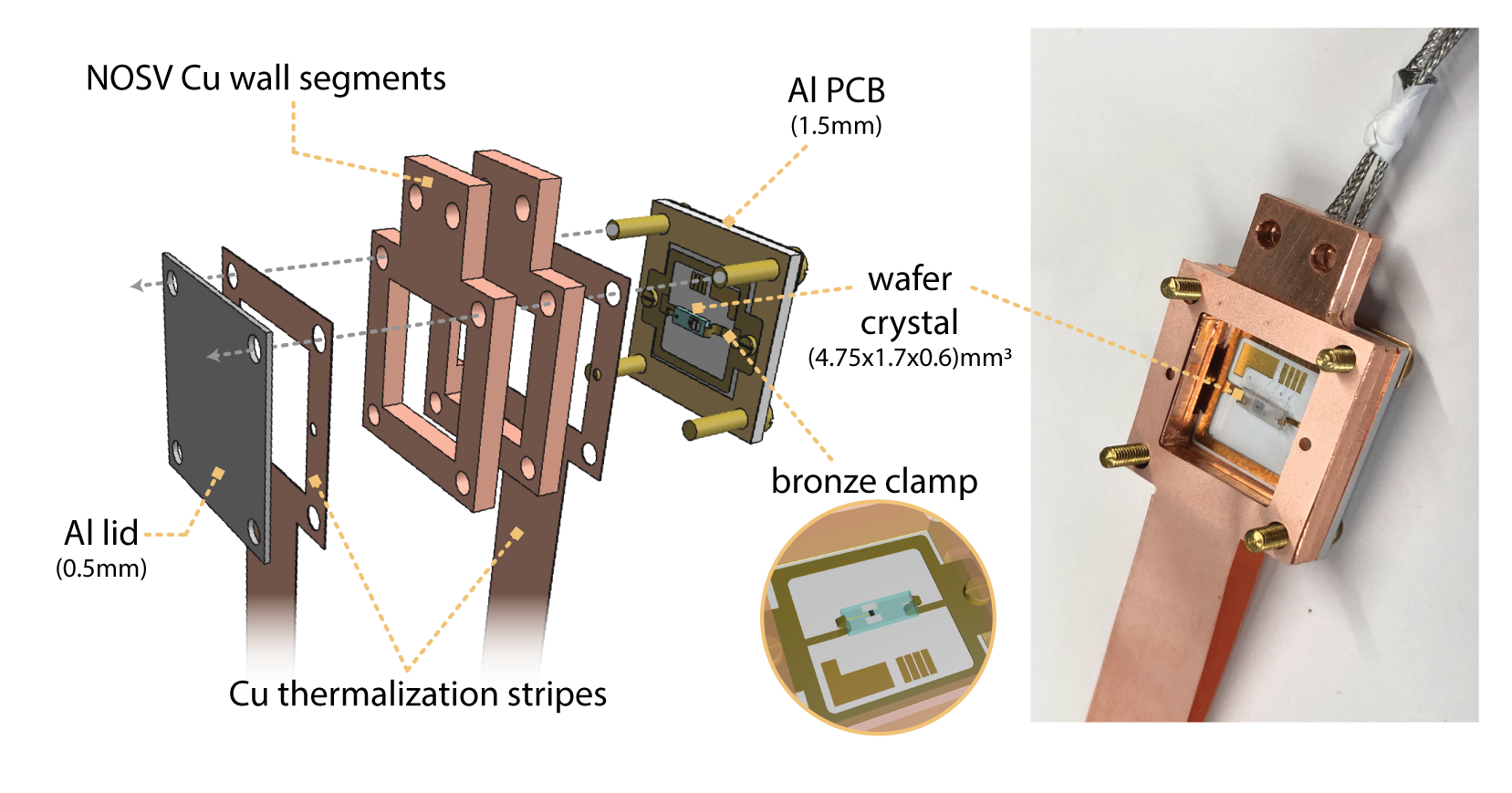}
    \caption{Exploded 3D view and picture of the detector holder, which consists of an aluminium-core PCB, a minimal housing made of a NOSV copper wall segment, a thin aluminium lid, and interleaved high-conductance copper stripes for direct thermal sinking of the PCB to the heat bath. The wafer is secured by flexible bronze clamping arms (zoom-in), whereas the cube (not shown) is pressed onto the PCB by a bronze clamp and thermally isolated by small aluminium spheres.}
    \label{fig:Crab_Dets}
\end{figure}

\begin{figure}[ht!]
    \centering
    \includegraphics[width=\linewidth]{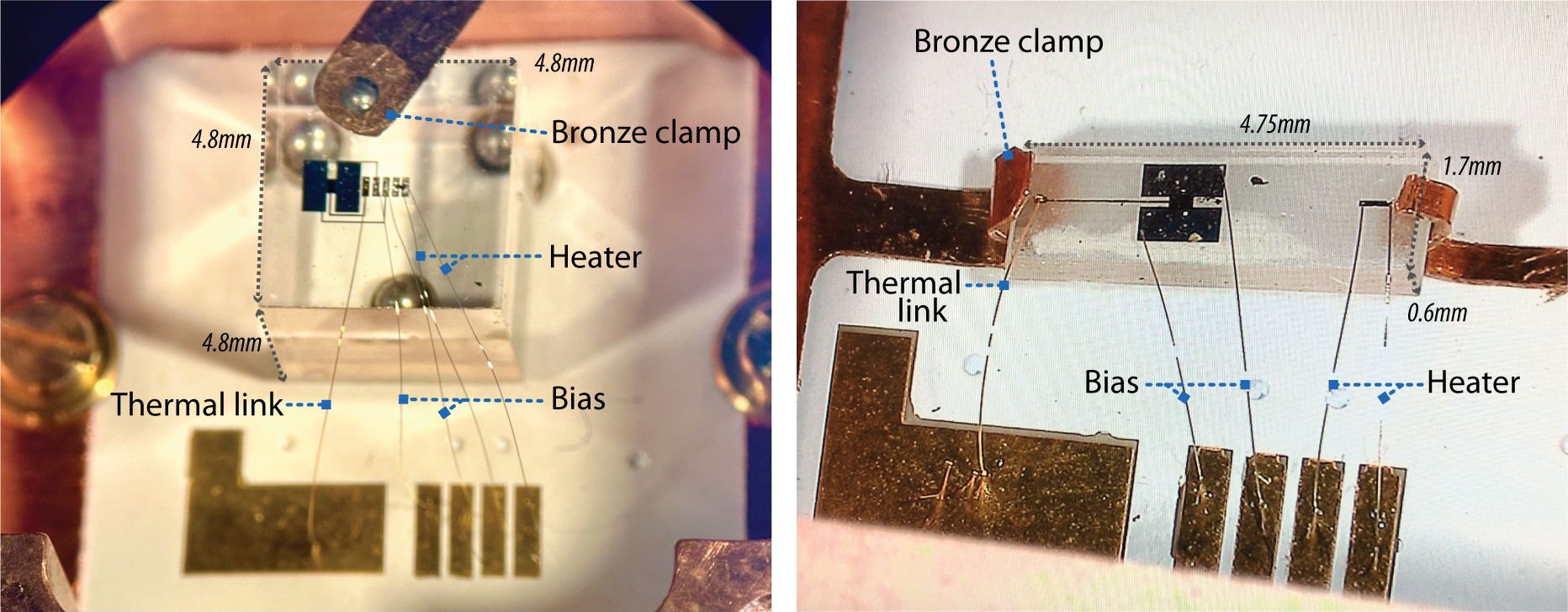}
    \caption{Pictures of two mounted and bonded crystals characterised for high-precision CRAB measurements: a cubic crystal (left) with dimensions of 4.8$\times$4.8$\times$4.8\,mm$^3$ and a wafer crystal (right) with dimensions of 4.75$\times$1.7$\times$0.6\,mm$^3$. Electrical contacts for the heater and bias lines are implemented using 25\,$\mu$m aluminium wire bonds, while thermal connection is provided by a 17\,$\mu$m gold wire bond. Bronze clamps with point-like contacts are used to hold the crystals securely in place.}
    \label{fig:Det_Geometry}
\end{figure}

We have developed dedicated ultra-low threshold cryogenic detectors for \CRAB, housed within an optimized low-mass mechanical support designed to minimize capture of stray neutrons in the surrounding material. The detector holder consists of an instrumented double-sided aluminium-core printed circuit board (PCB) (insulated metallic substrate) with plated through-holes, serving both as a structural support and an electrical interface. Aluminum was selected for its comparably low neutron capture cross-section. A minimal housing is incorporated to shield the detector from thermal radiation while keeping the overall mass low. To ensure efficient thermalization, interleaved high-conductance copper foils, thermally coupled to the PCB’s copper circuit layer, provide direct thermal sinking to the heat bath. To accommodate different absorber geometries, two modified versions of the holder have been developed: one for a cubic crystal and another for a significantly thinner, wafer, crystal (see Fig.~\ref{fig:Crab_Dets}). In the cubic configuration, a bronze clamp attached to the holder's wall segment presses the crystal onto the PCB. Small aluminum spheres, affixed in notches on the PCB and the clamp, ensure point-like contact while thermally isolating the crystal from the holder. For the wafer geometry, the crystal is secured by two flexible bronze clamping arms that apply gentle pressure from the side. Pictures of two mounted and bonded crystals of different geometries are shown in Fig.~\ref{fig:Det_Geometry}. 

Characterization measurements of the \CRAB detectors were conducted in the underground laboratory (UGL) of the Technical University of Munich (TUM) using the same wet $^3$He-$^4$He dilution refrigerator Kelvinox 100 from Oxford Instruments plc designated for \CRAB, just a few weeks before its relocation to the TRIGA Mark-II reactor site in Vienna. Data acquisition (DAQ) was performed using a hardware-triggered DAQ system, recording pulse traces with an Incaa Computers VD80 transient recorder at a sampling frequency of 10~kHz. Two CaWO$_4$ crystals of different geometries, along with one \AlO crystal, were mounted in the experimental volume of the cryostat and tested in two separate runs. Each absorber was supported by the low-mass \CRAB detector holder mechanics. The detectors were continuously exposed to a $^{55}$Fe X-ray source for energy calibration. The properties of the three detectors and the specifications of the measurements are listed in Table \ref{tab:det_prop}. 

\begin{table}[h!]
\caption{Overview of the specifications and benchmark performances of the selected detector substrates for \CRAB, measured at TUM UGL. All detectors achieved \CRAB-grade energy baseline resolutions below 6~eV.}
\centering
\renewcommand{\arraystretch}{1.5}
\setlength{\tabcolsep}{6pt}
\resizebox{\columnwidth}{!}{%
\begin{tabular}{l|ccc}
\hline
\textbf{Detector}          & \textbf{IPP27\_Ca-c1}         & \textbf{W1-682-9}          & \textbf{W1-710-5}         \\ \hline \hline
\textbf{Target material}          & \CaWO                   & \CaWO                 & \AlO           \\ 
\textbf{Dimensions (mm)}        & 4.75 $\times$ 1.7 $\times$ 0.6 & 4.8 $\times$ 4.8 $\times$ 4.8 & 5.0 $\times$ 5.0 $\times$ 5.0 \\ 
\textbf{Mass (g)}              &     0.03                          &                 0.67            &  0.50                           \\ 
\textbf{Transition temperature (mK)} &        14                       &              13          &    18                         \\ 
\textbf{Baseline resolution (eV)} &         3.8 $\pm$ 0.4~\text{\textit{(stat)}}\,$^{+0.0}_{-0.4}$~\text{\textit{(sys)}}     &  5.5\,$\pm$\,0.1         &                3.7\,$\pm$\,0.1                                    \\ \hline
\end{tabular}%
}
\label{tab:det_prop}
\end{table}

During benchmark performance measurements at TUM UGL, all three detectors selected for the \CRAB phase II demonstrated excellent performance with energy baseline resolutions below 6\,eV and corresponding energy thresholds below 30~eV, providing sufficient margin for high-precision measurements of all predicted (sub-)keV nuclear recoil peaks in \CaWO and \AlO following thermal neutron capture. By requiring coincidences in the external $\gamma$-detectors, these peaks can still be resolved even with degraded energy resolution of up to $\sim$\,10\,eV, but at the cost of reduced statistics.

\subsection{Cryogenic detector data acquisition system}

The new VDAQ3 (Versatile Data Acquisition 3) system, developed by the HEPHY electronics department, serves as the core data acquisition system for the \CRESST, \COSINUS and \NUCLEUS experiments using cryogenic detectors read out \textit{via} SQUIDs. Designed to replace earlier systems, VDAQ3 introduces key improvements including a modular architecture, digital control signal generation, scalable channel count (up to several hundred), optical data transfer, and continuous streaming with adjustable sampling rates. It offers enhanced frequency range and bandwidth, inter-module synchronization, and precise timing with external systems such as muon vetoes. VDAQ3 samples signals at 24-bit resolution and 1~MHz-sampling frequency, while TES operation data are transmitted at 18~bits and 1~MHz-sampling frequency. 

\subsection{Calibration systems for electronic recoils}

Comparing the calibrated electron-recoil energy scale to the nuclear-recoil calibration technique presented in this work is essential for understanding solid-state effects in the target crystals 
and enabling precision measurements of quenching factors at the lowest energies. Traditionally, radioactive X-ray sources -- primarily $^{55}$Fe emitting photons around 6~keV -- have been used to calibrate the electron-recoil response of TES-based cryogenic detectors. Extrapolation toward the lower-energy region relevant for dark matter and CE$\nu$NS searches is typically performed using artificial heater pulses, called test pulses, injected \textit{via} an ohmic heater on the detector. 
Recently, a novel LED calibration system \cite{Cardani_2018,DELCASTELLO2024169728} was integrated into the \CRAB setup at the TRIGA reactor in Vienna. Light pulses of variable intensity are delivered to the detectors \textit{via} optical fibers embedded in the cryogenic environment. With 255~nm wavelength the photons can travel deep in the \CaWO crystal and excite electrons in the conduction band. The de-excitation of these electrons is then read in the phonon channel. By analyzing the photon statistics and relative pulse amplitudes, this method enables absolute energy calibration across a broad energy range with a precise and continuous monitoring of detector response over time. The system is non-intrusive and can be deactivated during beam-on data taking, offering high flexibility.
In addition, the \NUCLEUS collaboration has developed an X-ray fluorescence (XRF) source that extends the calibration range down to sub-keV energies, providing discrete lines between 600~eV and 6~keV. This technique offers cross-validation and benchmarking of the LED-based calibration.

Employing multiple calibration methods for electron recoils is essential for the \CRAB project, as it enables a detailed understanding of low-energy phonon physics. Each technique probes different aspects of the detector response: LED pulses generate many low-energy optical photons distributed throughout the crystal volume, X-rays produce localized near-surface interactions, and artificial test pulses are injected form a discrete location close to the TES and differ in their primary phonon spectra. The interplay between these approaches provides complementary information and is key to fully exploiting \CRAB\!'s calibration capabilities. Finally, the response to LED pulses can also serve as a cross-calibration between the \CRAB facility in Vienna and the experimental site for CE$\nu$NS or DM measurements.

\section{Detection of high energy $\gamma$-rays}
\label{sec:gamma_det}

The detection of the high-energy $\gamma$-ray in coincidence with the nuclear recoil it has induced in the cryogenic detector is a unique feature of the \CRAB experiment installed at ATI. This technique significantly improves the signal-to-background ratio and relaxes energy resolution constraints for the detection of \CRAB calibration peaks, at the expense of a loss of statistics \cite{Thulliez_2021}. Our energy range of interest must cover the single-$\gamma$ de-excitations of the target nuclides after neutron-capture, typically in the 5--11~MeV range. As $\gamma$-detectors large \BaF crystals placed outside the cryostat were selected. \BaF is an inorganic scintillator offering good detection efficiency for multi-MeV $\gamma$-rays thanks to its density of 4.88~g/cm$^{\text{3}}$ \cite{annurev:/content/journals/10.1146/annurev-matsci-070616-124247}. The light emitted is shared between a fast component with a decay time of 0.6~ns and photons of around 200~nm wavelength, requiring a photomultiplier with a quartz window, and a slow component with a decay time of 630~ns and photons of 310~nm wavelength \cite{Laval1982,Marrone2006}. The fast component typically accounts for about 10--20\% of the total emitted light.

\subsection{BaF$_2$ detectors array }
\label{subsec:BaF2_detectionArray}

\begin{figure}[ht!]
    \centering
    \includegraphics[width=1.0\linewidth]{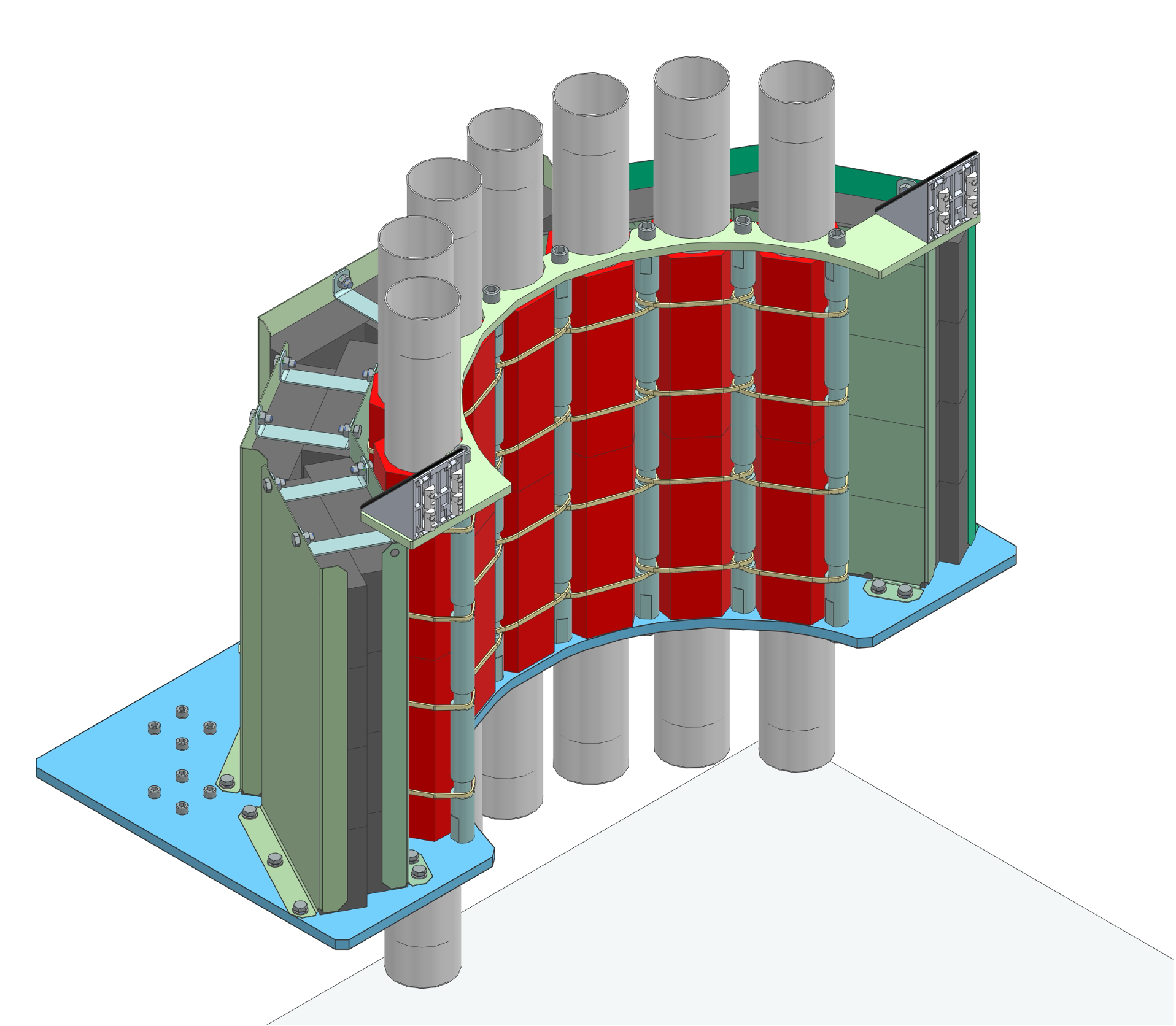}
    \caption{3D view of a half-crown of 14 $\gamma$-detectors, made of 7 vertical towers of two \BaF hexagonal crystals (red) read out by photomultipliers (grey cylinders) placed around the dewar. Two layers of 5~cm thick lead bricks on the outside of the detectors suppress the external $\gamma$-background. }
    \label{fig:Gamma_Detectors}
\end{figure}

A total of 40 \BaF crystals has been recovered from the “Château de Cristal” device, initially developed for nuclear spectroscopy in the 1980's \cite{Vivien1983,lopezjimenez2000,lima2004, bastin2007}. Each crystal has a hexagonal base with 50~mm sides, a height of 140~mm and is optically coupled to a photomultiplier with a quartz window. 28 of these detectors have been selected to build two half-crowns against the outer wall of the dewar as illustrated in Fig.~\ref{fig:Support_Structure} and \ref{fig:Gamma_Detectors}. \geant simulations performed with the \toucans code \cite{THULLIEZ2023} have shown that this arrangement of the crystals is the best compromise between solid angle coverage and intrinsic $\gamma$-detection efficiency. It is also quite compact around the deware allowing the implementation of lead shielding to protect the $\gamma$-detectors against the ambient background of high-energy $\gamma$-rays induced in the experimental hall when operating the reactor. 

\subsection{BaF$_2$ data acquisition system}
\label{subsec:BaF2_daq}

The \BaF signals are read by \faster electronic cards developed by LPCCaen~\cite{FASTER}. A $\mu$TCA crate houses one syroco\_amc\_c5 mother board and seven CARAS-CARAS daughter boards with four channels each. The charge integration QDC module of the CARAS-CARAS board is loaded onto the card's FPGA. Signals are digitized with 500~MHz sampling frequency by a 12-bit flash ADC in the $\pm$ 1.15~V range. 
A trigger is defined by a signal exceeding a threshold amplitude, set for each channel, for at least 28 ns. Only triggered channels are read out. The total charge Q$_{\text{total}}$ is computed in a [$-$30~ns;1500~ns] integration window around the trigger time while a fast charge Q$_{\text{fast}}$ is computed in [$-$30~ns;8~ns], providing Pulse Shape Discrimination capability.

\subsection{Characterisation of the BaF$_2$ detectors}
\label{subsec:BaF2_characteristics}

The energy response of the 28 detectors is characterized using known $\gamma$-lines from $^{\text{137}}$Cs (0.662~MeV), $^{\text{60}}$Co (1.173 and 1.332~MeV), Am-Be (4.440~MeV) radioactive sources as well as the 2.223~MeV line from n-H capture. The proportionality between the positions of the 5 detected peaks and their nominal energy is validated to better than 1\% over the entire energy range for all detectors. After tuning the high-voltage of the photomultipliers all gains are aligned on the mean value of 8.0~eV/QDC-channel, to within $\pm$2.5\%. The evolution of the standard deviation of the calibration peaks versus incident energy is well fitted in all detectors by the function
\begin{equation}
    \frac{\sigma(E)}{E} = \sqrt{\frac{a}{E^2}+\frac{b}{E}+c}\,. \\
\end{equation}
From these fits we find a mean relative energy resolution of 2.3\% with $\pm$0.2\% dispersion at 6.2~MeV, the energy of the single $\gamma$-transition associated to the main \CRAB recoil peak expected in \CaWO. Hence, the energy response of the \BaF detectors meets our initial specifications of a linearity at the few \% level and a resolution better than 5\% in the region of interest.

Full-peak detection efficiency was checked with a $^{\text{60}}$Co source whose activity is known to within 2.2\%, placed at a known distance from a \BaF crystal. This measurement revealed a full-peak efficiency surprisingly (23$\pm$3)\% below \geant's prediction. This result is confirmed by other studies \cite{lopezjimenez2000, lima2004}, which attribute this loss of efficiency to aging phenomena that have developed in these 40+ year-old detectors. Regeneration of the \BaF crystals by annealing \cite{Wei_1991,MA1993113} will possibly be considered to increase the statistics of coincidence events in future measurements. Combining this intrinsic detection efficiency with the total solid angle coverage of the detector crown, we estimate that the full peak detection efficiency at 7~MeV is 3.2\%. 

\begin{table}[ht!]
\caption{Parent nucleus and energy of the four main $\alpha$ lines identified in the \BaF counters. When setting the energy scale from the response to know $\gamma$-lines (see Sect. \ref{subsec:BaF2_characteristics}), the response to $\alpha$ particles appears as significantly quenched. The mean detected energy is reported in the third line of this table, however $\sim$10\% dispersion of the quenching factor is observed among the counters.}
\label{tab:alpha}
\begin{tabular*}{\columnwidth}{@{\extracolsep{\fill}}r|rrrr@{}}
\hline
Nuclide & $^{226}$Ra & $^{222}$Rn & $^{218}$Po & $^{214}$Po \\
\hline\hline
E$_{\text{emitted}}$ (MeV) & 4.87 & 5.59 & 6.11 & 7.83 \\
E$_{\text{detected}}$ (MeV) & 1.47 & 1.75 & 2.04 & 2.86
 
\end{tabular*}
\end{table}

Because of a chemical affinity between barium and radium, \BaF crystals are contaminated with $^{226}$Ra \cite{Belli_2014}. The associated $\alpha$-decay chain leads to 4 main lines listed in Table \ref{tab:alpha}. They induce a total rate of $\approx$195 Hz above 1 MeV on each counter (see Fig.~\ref{fig:Gamma_Background}). Due to the quenching effect of $\alpha$-induced scintillation compared to $\gamma$($\beta$) interaction of the same energy, they are all reconstructed below 3~MeV $\gamma$-equivalent, \textit{i.e.} below the region of interest of the single-$\gamma$ transitions induced by neutron capture. No accurate absolute calibration can be inferred from these $\alpha$-peaks because the quenching factor varies within $\pm$10\% from one detector to the other. However, they can be used for an online monitoring of the response stability of each detector. Since the fast component is suppressed in the case of  $\alpha$ particles, the Q$_{\text{fast}}$/Q$_{\text{tot}}$ ratio can also be used to discriminate efficiently $\alpha$ from $\gamma$ signals in the few MeV energy range.  

\section{Simulation package}
\label{subsec:simulation}

The \CRAB experiment is accurately simulated with the \toucans code~\cite{THULLIEZ2023} based on \geant version 11.3~\cite{Geant42003,Geant42006,Geant42016}. The description of the secondary neutron beam sent to the cryostat includes its measured energy spectrum, angular divergence and transverse extension. The geometry of the various parts is implemented in detail, including the successive beam collimators, the crowns of $\gamma$-detectors with their shielding, the structure of the dewar, the cryostat and the cryogenic detectors in their holder. Neutron transport through all materials is performed with the \geant Neutron-HP package. Recent developments~\cite{Mendoza2018,Tran2018a,Thulliez2022,Zmeskal2023,Zmeskal2024} have brought version 11.3 up to the same level of accuracy as reference codes such as \mcnp~\cite{Werner2018} or \tripoli~\cite{Brun2015}. Low-energy electromagnetic interactions rely on Livermore physics list when precise description of photon propagation with energy of the order 100~eV is required and EMZ physics list otherwise.
\newline 
\indent
Once a thermal neutron is captured by a nucleus of the cryogenic detector, the core de-excitation process \textit{via} emission of $\gamma$-rays and conversion electrons and the induced nuclear recoil in the detector are described by the \fifradina code \cite{Soum2023}. In this package the output of the \fifrelin nuclear de-excitation code~\cite{Litaize2015,FIFRELIN_userGuide}, predicting the energy and emission time of the particles ($\gamma$, electron and positron) in the de-excitation cascade, is coupled to the simulation of the displacement cascade by the \iradina code~\cite{Borschel2011}. To our knowledge, this approach offers the best treatment of interleaved time developments of nuclear de-excitation and displacement cascades. While it has no impact on the mono-energetic recoils induced by single-$\gamma$ de-excitations, it is crucial for an accurate prediction of the recoil spectrum induced by multi-$\gamma$ cascades. In particular, it can happen that the target nucleus that started to recoil after the emission of the first $\gamma$-ray has time to stop in matter before the emission of the next $\gamma$-ray. In that case the detected recoil is again mono-energetic
\begin{equation}
\label{eq:slow_hypothesis}
E_\text{r} = \sum_i E_{\text{r}_i} = \sum_i \left| \vec{P_{\gamma_i}} \right|^2\ / \ 2M_\text{n}    
\end{equation} 
with $\vec{P_{\gamma_i}}$ the momentum vector of the $i^\text{th}$ $\gamma$-ray and $M_{\text{n}}$ the mass of the recoiling nucleus. This feature potentially produces extra calibration peaks, whereas neglecting the lifetime of the nuclear exited levels leads to a continuous distribution of recoils sensitive to the relative orientation of the emitted $\gamma$-rays with 
\begin{equation}
\label{eq:fast_hypothesis}
E_\text{r} = \left|\sum_i \vec{P_{\gamma_i}} \right|^2\ / \ 2M_\text{n} \,. 
\end{equation}
\fifradina predictions for most materials of interest for cryogenic detectors have been made available to the community in an open access repository~\cite{soum_sidikov_2023_7936552} along with a \geant interface named \fifrelingeant~\cite{thulliez_loic_2023_7933381} for direct use of them when simulating thermal neutron radiative captures.

\section{Commissioning data}
\label{sec:commissioning}

Few months after its installation on the reactor site, the cryostat has been operated with good stability over several weeks. However, energy resolution has not yet stabilized at a level sufficient to access the calibration peaks in the 100~eV range expected from neutron captures in \CaWO. This point is currently being addressed by upgrading the cryogenic detector electronic chain and grounding scheme. In the meantime, we report on the analysis of complementary data that provide crucial proofs of concept for future high-precision measurements in the 100~eV range. We first study the energy response of the $\gamma$-detectors and the mitigation of the high-energy background induced by the reactor operation, in view of the detection of events in coincidence with the cryogenic detector. We show that the \BaF counters can also be used to build an accurate model of the $\gamma$-background induced in the cryogenic detectors by the reactor activity. Then rate and spectrum analyses of high-energy events, on a 1 to 100~keV scale, in the one cryogenic detector operated so far are presented. We demonstrate a very good understanding of beam induced events as well as external background.

Finally, we establish for the first time the tagging of neutron capture events by the detection of a high-energy $\gamma$-ray outside the cryostat, in coincidence with an energy deposition in the cryogenic detector. 

\subsection{Gamma detector data}
\label{subsec:gamma_data}

\begin{figure}[ht!]
    \centering
    \includegraphics[width=1.0\linewidth]{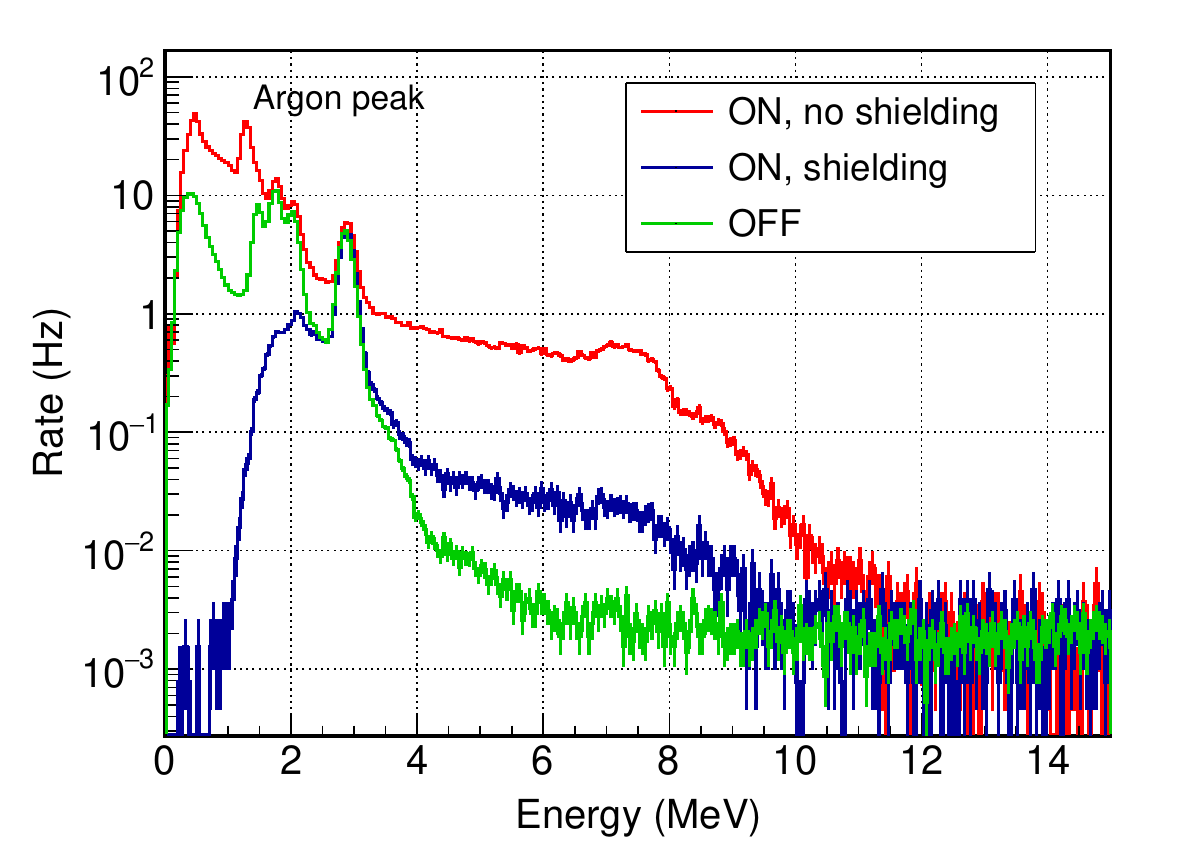}
    \caption{Typical energy spectra measured in one of the \BaF detectors taken with the reactor stopped (green), or operating at full power, with (blue) and without (red) shielding around the detectors. In standard running conditions the detection threshold is set just below the last $\alpha$ peak as seen in the blue spectrum.}
    \label{fig:Gamma_Background}
\end{figure}

Figure~\ref{fig:Gamma_Background} shows typical energy spectra from one of the \BaF detectors. In the low-energy region (E $\leq$ 3.5~MeV) the rate is dominated by $\alpha$ contamination, with four peaks clearly visible. Their energies are reported in Table \ref{tab:alpha}. In standard running conditions, the detection threshold is set slightly below 2~MeV to reduce the total counting rate per counter while preserving the $\alpha$-peak with the highest energy (2.86~MeV), for accurate monitoring of the detector response. The position of this peak can be measured with percent statistical precision every few minutes and a stability of 0.5\% is observed over a 9-day period. 

In the region of interest for tagging the high-energy $\gamma$-rays from \CRAB events, 4.7 to 7.4~MeV for a \CaWO target, the count rate is 0.3~s$^{-1}$ per counter when the reactor is off and dominated by the cosmic rays contribution. At full reactor power, the rate increases by approximately two orders of magnitude up to $\approx$40~s$^{-1}$ per counter. Assuming a coincidence time window of 500~$\mu$s, based on the rise time of the cryogenic detectors, this reactor-induced background would correspond to an unacceptable probability of accidental coincidence of 56\%. 

The spectral features in the 7 to 8~MeV range point to high-energy $\gamma$-rays induced by the capture of the ambient thermal neutrons on nearby items, and particularly on iron, as confirmed by measurements performed with a portable high-resolution Ge detector. An extensive test campaign has shown that several source terms were responsible for this background leading to the implementation of a multi-component complementary shielding, on top of the ring of lead bricks illustrated in Fig.~\ref{fig:Gamma_Detectors}: the entire cryostat support structure has been covered with boron-doped rubber mats to block thermal neutrons and prevent radiative neutron captures too close to the detectors; a stronger entrance collimator has been installed around the beam axis with a first aperture of \diameter~=~2~cm through a Boron-loaded screen followed by 5~cm of lead and an second square aperture of $1.1\times1.1$~cm$^{2}$ installed against the dewar wall; an additional 10~cm thick lead screen has been mounted on a mobile cart to mask the solid angle of the reactor wall; finally, the floor underneath the cryostat has been covered with 4~cm of lead. All these components gradually reduced the external background to the final spectrum shown in Fig.~\ref{fig:Gamma_Background} (blue histogram), with a mean rate of 2.3~s$^{-1}$ per counter in 4.7 to 7.4~MeV range, leading to acceptable probability of accidental coincidences.

\begin{figure}[ht!]
    \centering
    \includegraphics[width=1.0\linewidth]{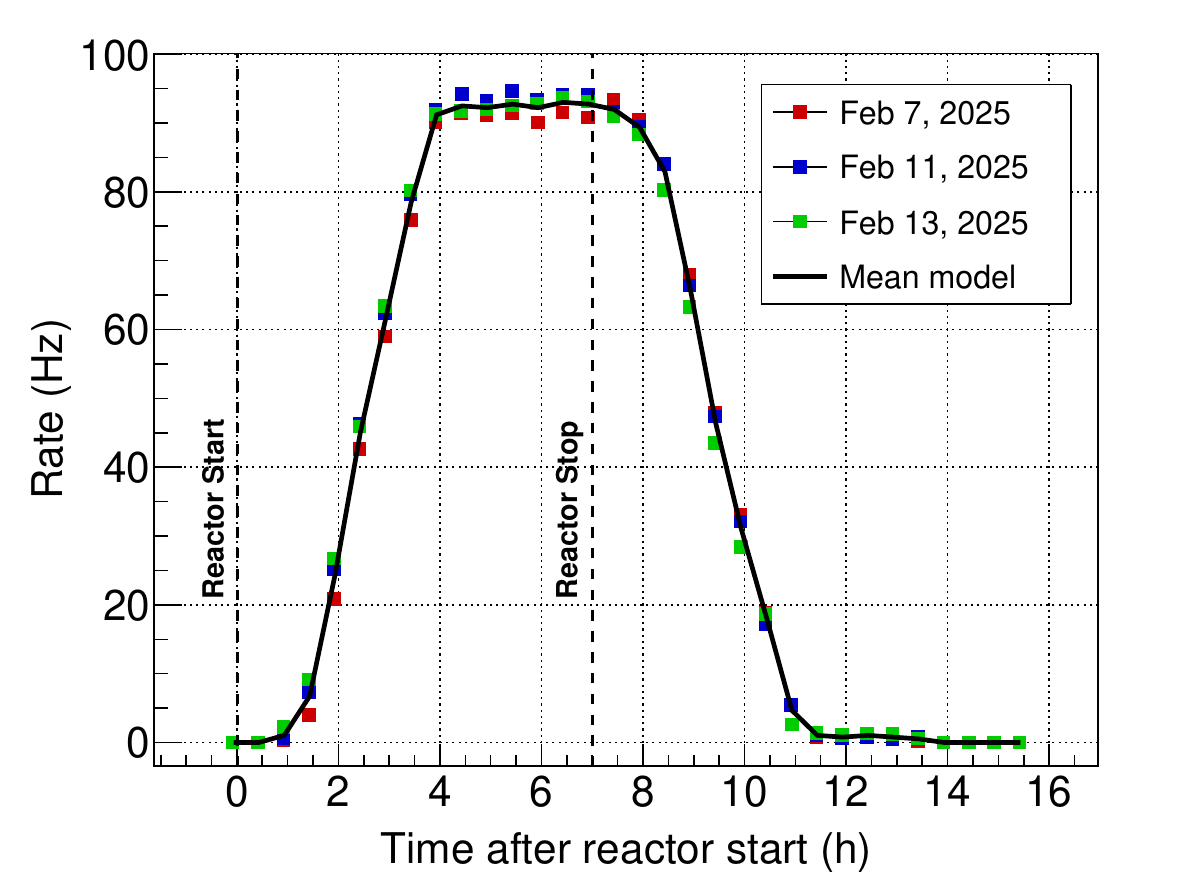}
    \caption{Activity of $^{41}$Ar (E$_\gamma$ = 1.29~MeV) measured in the external \BaF detector for three different days of reactor operation at full power (colored points). For each day, the time origin is taken at reactor start. The model of $^{41}$Ar activity is defined as the average of the three measurements (black curve).}
    \label{fig:Model_41Ar}
\end{figure}

When the reactor is running, we observe the growth of a peak centered at an energy of 1.29~MeV, indicating neutron activation of $^{41}$Ar from air. One can see in Fig.~\ref{fig:Gamma_Background} that this peak is actually intense and could dominate the cryogenic detectors background induced by external $\gamma$-rays. Therefore, an additional \BaF counter was placed outside the lead crown, but still inside the cryostat rack, to monitor the time evolution of this background. A reproducible pattern of activation and decay is observed over three days of nominal reactor operation with full power for seven hours per day (see Fig.~\ref{fig:Model_41Ar}). The mean values define the shape of a ``\,$^{41}$Ar model\," used in the rate-only analysis presented in Sect. \ref{subsec:rate_only}. The main source is identified as the volume of air contained in a pneumatic rabbit system running through the reactor core to circulate irradiated samples. The air close to the core is activated by the very high neutron flux and then slowly diffuses to the rest of the tubing running in the reactor hall in the area of our experiment. This explains why the rise and decay of the measured activity appear to be delayed by about an hour with respect to reactor start and stop times, with significant deviation in shape from a naive exponential growth and decay driven by the 110~min half-life of $^{41}$Ar.

\subsection{Cryostat and detector performance}
\label{subsec:performance}

This section summarizes the cryogenic run history at the reactor site and benchmarks the achieved cryostat and detector performance, focusing on the baseline energy resolution and detector stability. To date, three cryogenic runs have been performed with the two \CaWO detectors IPP27\_Ca-c1 and W1-682-9 (wafer and cube geometry, respectively) and aligned with the neutron beam. In all three runs, the cryostat reached a base temperature of below 10\,mK and both W-TES went into superconducting transition. The commissioning data show good detector stability over timescales of several days to weeks. Notably, over a duration of a few hours during Run 1, the \CaWO detector IPP27\_Ca-c1 reached a promising energy baseline resolution of $\approx$\,7\,eV --- only a factor less than 2 above the targeted benchmark performance demonstrated in Sect. \ref{sec:det_benchmark}. However, the electronic noise situation after integration of the setup at the new experimental site and the commissioning of the new VDAQ3 system is not yet under control, preventing the energy resolution from stabilising at a level sufficient to access the calibration peaks in the 100\,eV range expected from neutron captures in a \CaWO target. Efforts to mitigate this issue are ongoing, including an upgrade of the cryogenic detector electronics chain and a revision of the grounding scheme and electrical layout.

\begin{figure}[h!]
    \centering
    \includegraphics[width=1\linewidth]{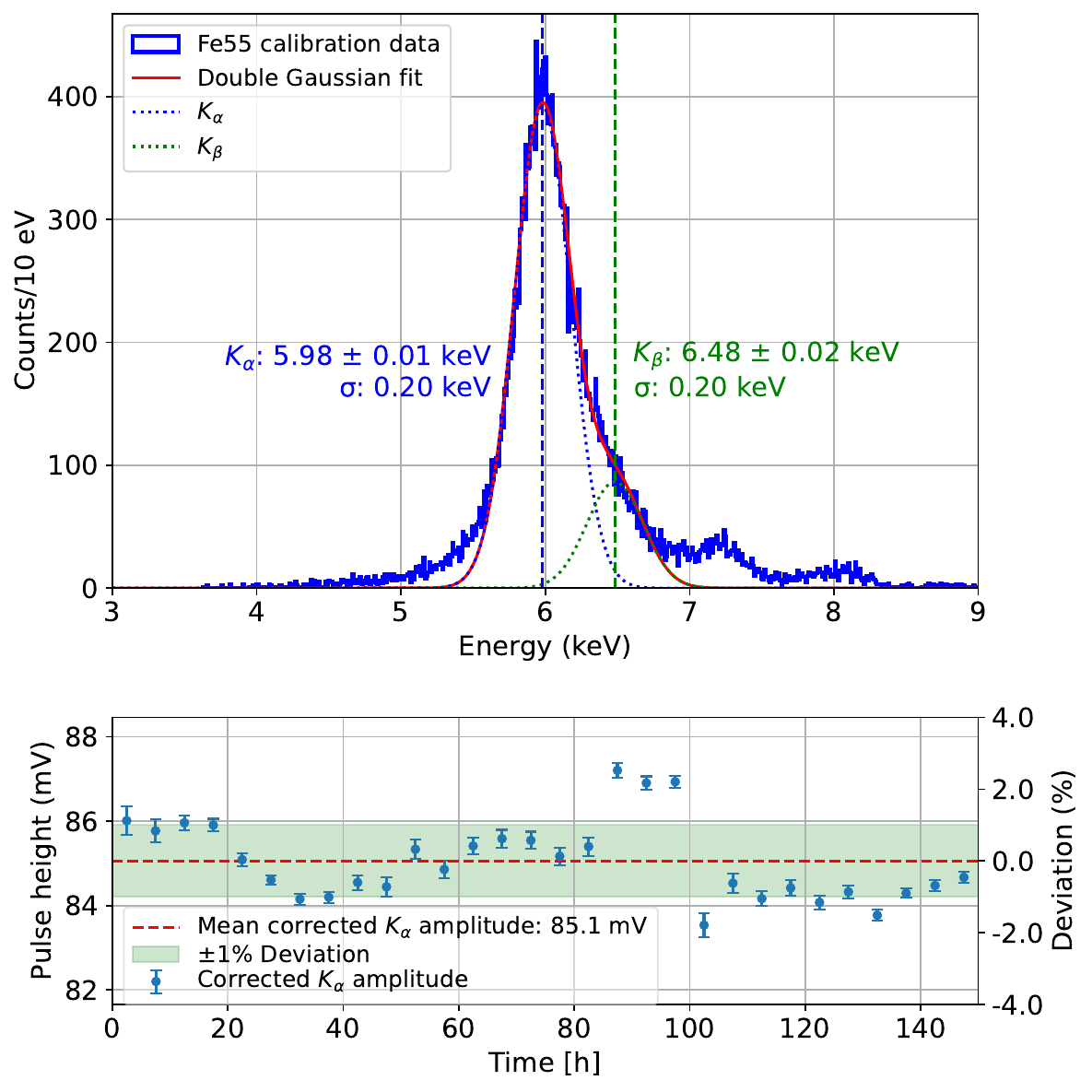}
    \caption{$^{55}$Fe calibration spectrum and stability of particle pulses during a 150~h background measurement with the \CaWO wafer detector IPP27\_Ca-c1 in Run 2. 
    For the particle pulses, we select 5.9~keV events corresponding to the K$_{\alpha}$ line induced by the $^{55}$Fe source. The strong correlation between the stability of heater and particle pulses is used to correct for a $\pm$\,5\% variation in detector response over time. The resulting stability of the K$_{\alpha}$ line is at the percent level. One possible explanation for the spectral features at higher energies are direct energy depositions in the TES. They are accounted for by a conservative asymmetric systematic uncertainty in the calibration.}
    \label{fig:det_stability_run2}
\end{figure}

\begin{figure}[h!]
    \centering
    \includegraphics[width=1\linewidth]{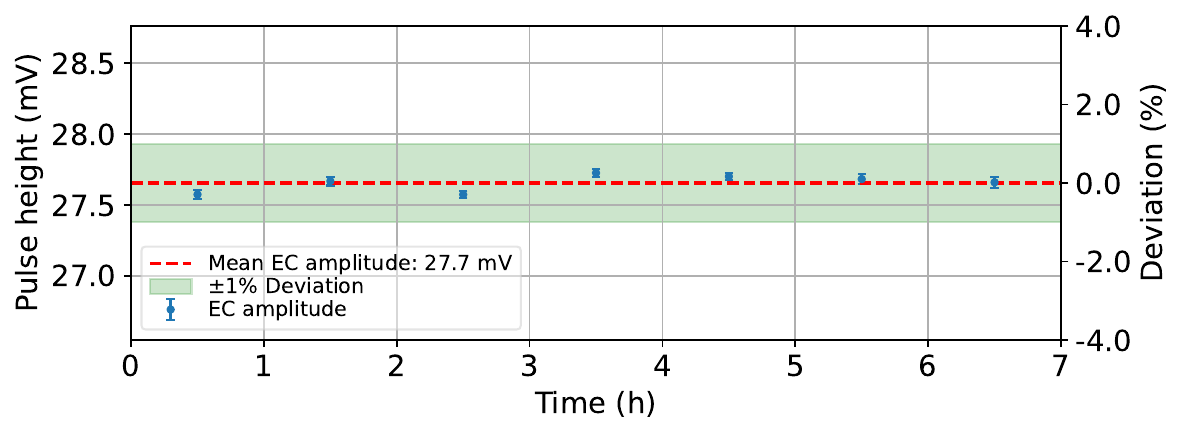}
    \caption{Stability of particle pulses during a 7\,h beam-ON measurement with the cubic \CaWO detector W1-682-9 in Run 3. We select 47~keV events induced by conversion electrons following neutron capture (see Fig.~\ref{fig:energy_spectrum}). The particle pulse height remains stable within $<$ 1\% over time.}
    \label{fig:det_stability_run3}
\end{figure}

In the following a brief summary of the cryogenic run history at the reactor site is presented:
\begin{itemize}
    \item \textbf{Run\,1} (Sept. 4, 2024 – Oct. 6, 2024): First cryogenic run after setup assembly at the reactor site. A vacuum leak in the cryostat-external section of the mixing circuit prevented stable cryostat performance and terminated the run with a forced warm-up. First efforts were dedicated to mitigating electronic noise in the detector system. During an intermediate configuration, we observed the best onsite detector performance to date for several hours on the \CaWO detector IPP27\_Ca-c1, achieving a baseline energy resolution of (7.35\,$\pm$\,0.17)~eV.
    \item \textbf{Run\,2} (5 Nov. 2024 – 19 Dec. 2024): After fixing the vacuum leak issue, stable cryostat performance was achieved over several weeks. We conducted a first measurement with the \CaWO detector IPP27\_Ca-c1 with active detector stabilization \textit{via} PID and collected 150\,h of stable background data, however, with degraded energy resolution of (20.9 $\pm$ 0.2~\textit{(stat)}\,$^{+0.0}_{-3.8}$~\textit{(sys)})~eV. Fig.~\ref{fig:det_stability_run2} shows the $^{55}$Fe calibration spectrum of the background measurement and the stability of particle pulses over time. The detector's very slow response time and unstable noise conditions degrade baseline restoration and limit the efficiency of the PID control. As a result, we observe a $\pm$\,5\% variation in heater pulse height over time. In the analysis, we use the strong correlation between the stability of heater and particle pulses to correct for these instabilities. As shown for the stability of iron events over time, this correction restores particle pulse stability to the percent level. The dominant calibration line of the $^{55}$Fe source at 5.98~keV (Mn K$_{\alpha}$) is used to calibrate the pulse spectrum. The Mn K$_{\beta}$ line is identified at (6.48\,$\pm$\,0.02)~keV by a double-Gaussian fit, in good agreement with the expected literature value. Events corresponding to additional spectral features observed at higher energies (7--8~keV) exhibit distinct pulse shapes compared to the events of the main distribution, possibly indicating direct energy depositions of $^{55}$Fe gammas in the TES. To account for potential misidentification of the Mn K$_{\alpha}$ line, we assign a conservative one-sided systematic uncertainty, based on a relative $\approx$\,18\% shift in the calibration factor when evaluated on the higher-energy features. This shift is propagated to the reported asymmetric systematic uncertainty on the baseline energy resolution. The \CaWO detector W1-682-9 suffered from electronic noise and was not operated long-term.
    \item \textbf{Run\,3} (Jan. 23, 2025 – Feb. 14, 2025):  The \CaWO detector W1-682-9 was unintentionally operated at low sensitivity in a high dynamic range mode (presumably due to a parasitic series resistor in the detector readout circuit) and showed sub-percent level stability even without active detector stabilization. Fig.~\ref{fig:det_stability_run3} illustrates this stability of particle pulses over time during a 7~h beam-ON measurement. The low gain together with the good stability provided ideal conditions for studying the response to high-energy events associated with the neutron beam (presented in Sect. \ref{subsec:rate_only} and \ref{subsec:spectrum}). The \CaWO detector IPP27\_Ca-c1 showed similar performance to Run 2.
\end{itemize}

The analysis of the cryogenic detectors is performed using two independent frameworks: CAIT~\cite{CAIT} and Cryolab~\cite{XRF_paper} and is based in both cases on an optimum filter (OF) approach. This approach, using a frequency filter built from real particle and noise traces, is commonly used in cryogenic experiments as it maximises the signal-to-background ratio in amplitude estimation \cite{Di-Domizio_2011}. Data streams recorded by VDAQ3 are continuously sampled at 10\,kHz and contain both particle pulses and artificially injected heater pulses of known periodicity and energy, used to monitor and stabilize the detector response. Triggering is performed offline by identifying timestamps where the data stream exceeds a defined threshold. Before data processing, average pulse template and noise power spectrum are constructed from selected particle and noise traces for the optimum filter calculation. Pulse amplitudes are estimated from the maximum of the filtered trace, which, within the detector's linear range, corresponds to the event energy. For saturated high-energy events, a truncated template fit \cite{Felix_Wagner_thesis} is applied, restricting the fit to the linear region and extrapolating the amplitude. In addition to the trigger timestamps and pulse amplitudes, several pulse shape parameters --- such as pulse rise time, decay time, pre-trigger baseline --- are calculated to identify valid pulses and reject artifacts. The baseline energy resolution is estimated by applying the optimum filter to a large set of noise traces and sampling the noise amplitude at a fixed position. The resulting Gaussian distribution, centered at zero, provides the resolution estimate \textit{via} its standard deviation.

\subsection{Rate only analysis}
\label{subsec:rate_only}

The larger volume of the \CaWO cubic detector W1-682-9 offers greater statistical precision without being penalized by too large pile-up effects, thanks to a sufficiently fast response time of 80 ms. Its unintentionally low gain made it a unique laboratory for studying the high-energy counterpart of neutron captures. In fact the \CRAB method is based on neutron capture events in which all the particles in the nuclear de-excitation cascade escape from the cryogenic detector, and nuclear recoil is the only energy deposited. However, in most cases, an energy deposit from a conversion electron or from a low-energy $\gamma$-ray, typically between few 10~keV and few MeV, is detected in addition to the nuclear recoil. To less extent, part of the de-excitation cascade from neutron captures on the mechanics directly in front of the detectors can also reach the crystal. These large-amplitude events have the advantage of being relatively insensitive to baseline resolution, while allowing us to test the characteristics of the neutron beam and its interactions with the crystal and its environment. 

The trigger times are determined by differentiation of the data stream sample by sample after downsampling by a factor of ten (1 kHz frequency) and a high-energy event is tagged as a difference of at least $+$4~mV between consecutive samples. This differentiated stream approach is robust and ensures a nearly 100\% trigger efficiency, requiring no further correction of the measured rates.  

\begin{figure}
    \centering
    \includegraphics[width=\linewidth]{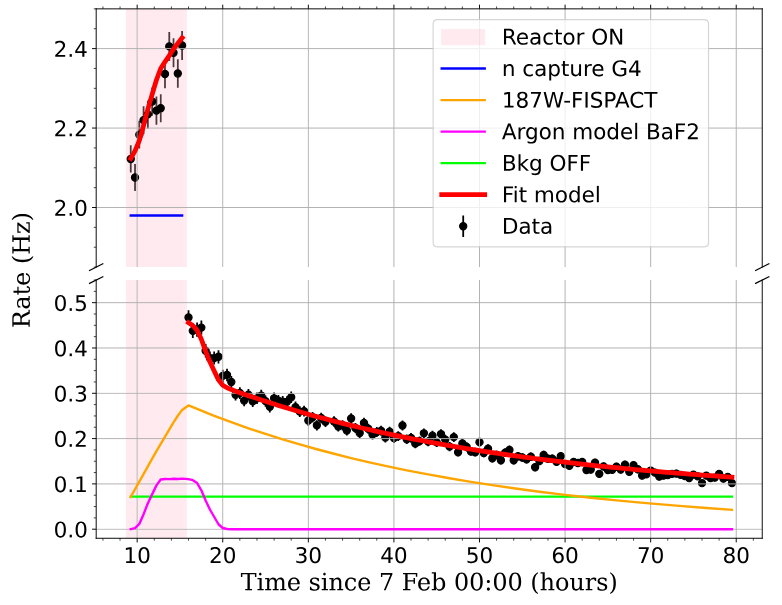}
    \caption{The black points show the time evolution of the measured counting rate above 20~keV during 7-hour shift of reactor operation at full power (shaded area) followed by a 3-day period with stopped reactor. A model with only 3 free parameters fits very well the data (red curve). The other colored lines illustrate the 4 components of this model: the simulated beam-induced rate with reactor at full power (blue), the simulated activation and decay of $^{187}$W (orange), the $^{41}$Ar model (magenta) and the constant background from ambient $\gamma$-rays when the reactor is stopped(green curve). See text for details on the fit parameters.}
    \label{fig:Rate_Analysis}
\end{figure}

Figure~\ref{fig:Rate_Analysis} shows the evolution in time of the measured rate, starting with one reactor shift (7~h) at full power, followed by 3 days without reactor activity. 
For an accurate comparison with predictions the amplitude in mV of all events triggered by the bin-to-bin analysis is determined with the OF analysis presented in the previous section. 
A software threshold of 15~mV is applied corresponding to 20~keV energy according to the calibration curve of Fig.~\ref{fig:calib_curve}. This threshold corresponds to when the 100 \% plateau trigger efficiency of the differential trigger is reached with 4 mV threshold. A model consisting of four main contributions is fitted to these data with only three free parameters. The beam-induced event rate due to neutron capture in the \CaWO crystal or surrounding materials is predicted by the \toucans simulation. The capture of neutrons in the crystal also produces instable $^{187}$W nuclides that decay by $\beta$-radioactivity with a lifetime of 23.8~h. This adds up to the prompt de-excitation cascades as a delayed beam-induced rate, driven by the history of reactor power and simulated with the FISPACT code~\cite{SUBLET201777}. These first two contributions scale with the neutron flux $\phi_\text{n}$, the first free parameter of the fit. In addition to the beam-induced rate, the model contains two background components: the $^{41}$Ar activation taken from Fig.~\ref{fig:Model_41Ar} and rescaled by the second fit parameter $K_\text{Ar}$, and a constant rate $R_\text{off}$, as third fit parameter that describes the ambient radioactivity, uncorrelated with reactor operation. We checked that, with the reactor always running at full power, the count rate is an affine function of beam intensity we send to our cryostat, and that the intercept at zero intensity was compatible with the contributions from argon and $R_\text{off}$, with no measurable effect from any additional contribution.
This model is in remarkable agreement with the data. The best-fit has a $\chi^2$ of 150.4 for 137 degrees of freedom (p-value = 0.20) and the parameter values are $\phi_\text{n}$ = (441.8 $\pm$ 1.7)~cm$^{-2}$s$^{-1}$, compatible with the direct flux measurement of Sect. \ref{subsec:crab_line}, $K_\text{Ar}=(1.25\pm 0.06) \times 10^{-3}$, of the order of the surface ratio of the \BaF and cryogenic detectors and $R_\text{off}=(72.6\pm 0.8)\times$10$^{-3}$~s$^{-1}$. 

\subsection{Spectrum analysis}
\label{subsec:spectrum}

An energy spectrum analysis of the same W1-682-9 cubic detector data is performed using each trigger identified in the differentiated stream analysis to save the trace at full sampling frequency in a time window extending from $-$256 samples to +768 samples around the reference trigger time. Then the pulse amplitude is determined more accurately with the OF presented in Sect. \ref{subsec:performance}. 
With this approach, cuts must be applied to the data to remove pile-up effects that could lead to incorrect energy reconstruction. Specifically, we require a flat baseline before the trigger (first 128 samples of the record length) and that only a single pulse is present in the trace. The cut efficiency was determined by simulating pulses with known amplitudes superimposed on noise traces randomly sampled from the data stream, accurately reproducing pile-up conditions. The estimated cut efficiency is $0.71\pm0.01$ and remains constant within the energy range of 10 to 230~keV.

\begin{figure}
    \centering
    \includegraphics[width=\linewidth]{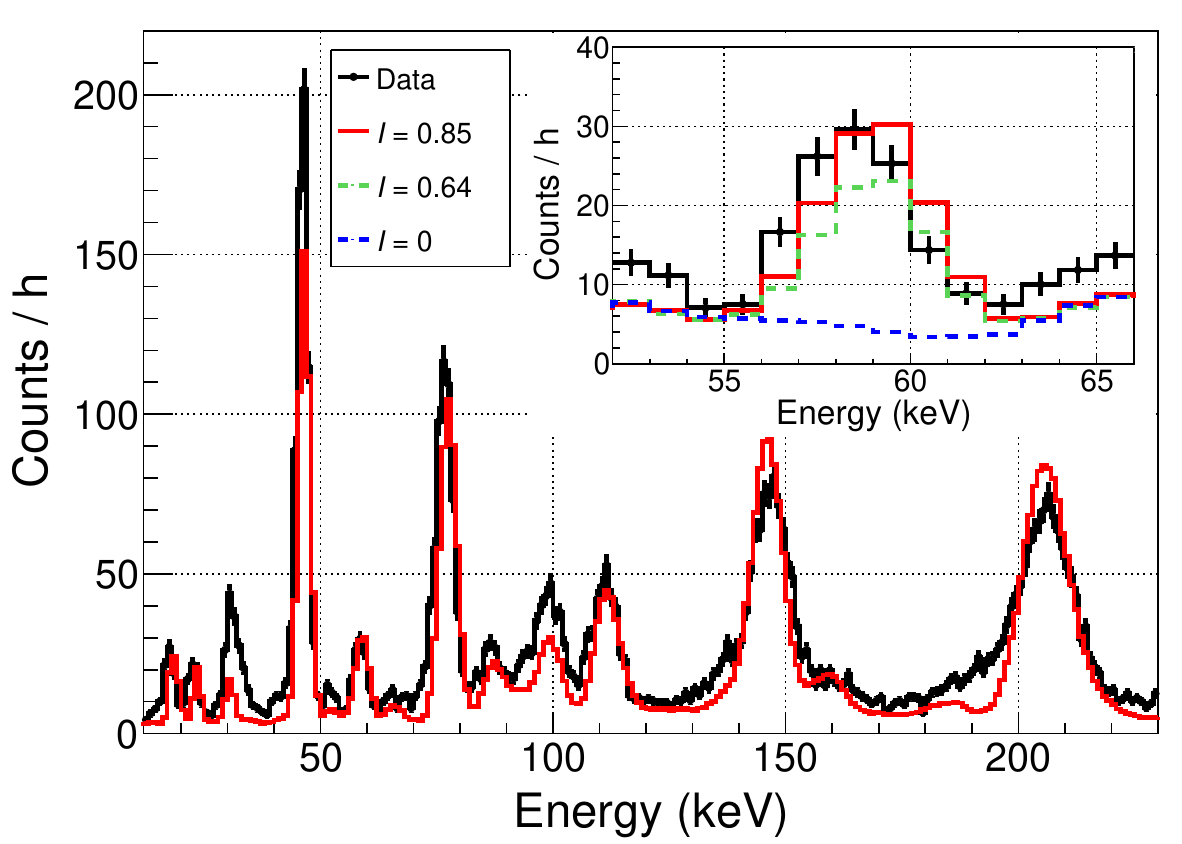}
    \caption{Energy spectrum of beam-induced events measured in the \CaWO cubic W1-682-9 detector (black points) compared to the prediction (red curve). The inset zooms in on the 59~keV peak initially missing in the simulation (dashed blue curve). The updated decay scheme of $^{187}$W (see Fig.~\ref{fig:decay_scheme}) allows to describe well this peak (red curve) while decay scheme with lower relative intensity $I$ of the transition, taken from \cite{PhysRevC.89.014606}, shows a deficit with respect to data (green curve).}
    \label{fig:energy_spectrum}
\end{figure}

\begin{figure}
    \centering
    \includegraphics[width=\linewidth]{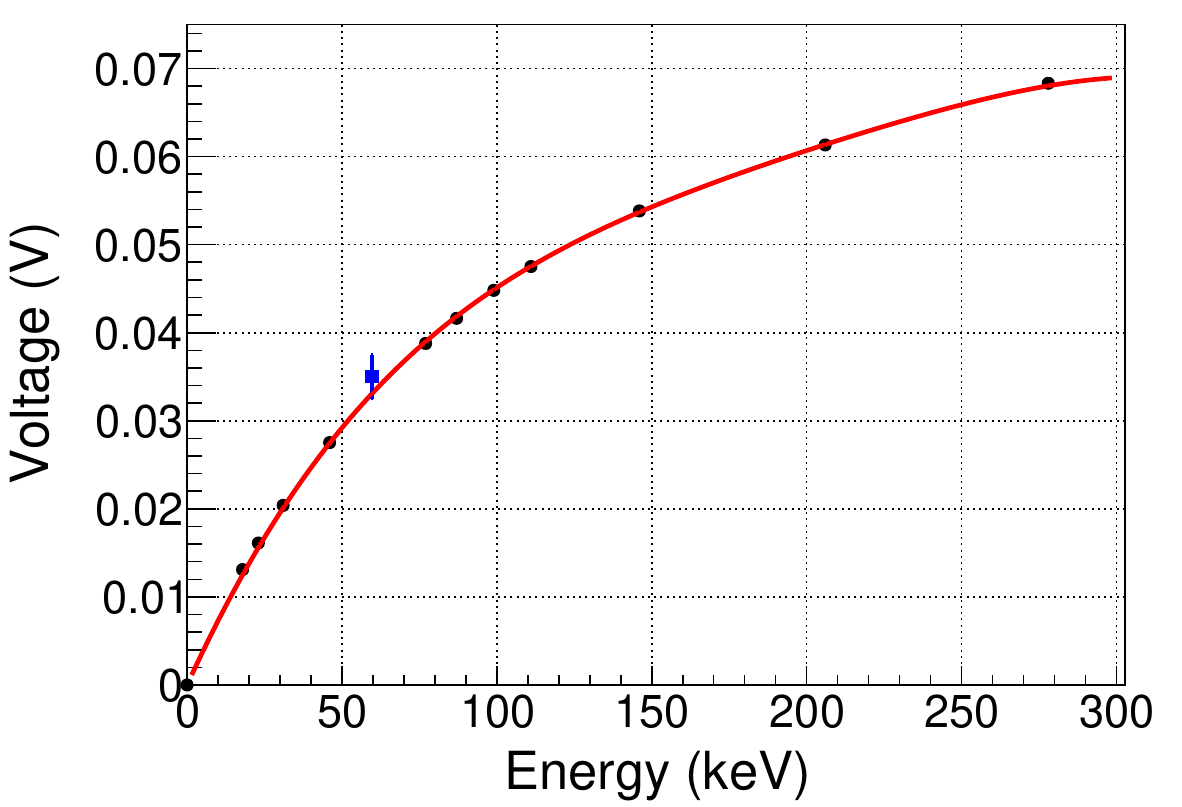}
    \caption{Correspondence between the measured position of the 11 most intense peaks in volts and their energy as predicted by the simulation (black points). A 4\textsuperscript{th} degree polynomial with the intercept set to zero is used as an effective calibration curve fitting all points with \% precision (red curve). An independent calibration point from the 59.5~keV $\gamma$-ray line of an $^{241}$Am source (blue point) is found in very good agreement with this curve.}
    \label{fig:calib_curve}
\end{figure}

The energy spectrum is corrected for this cut efficiency and background subtracted. The background spectrum is estimated from the reactor-OFF data taken in a period of time with a decay rate of $^{187}$W similar to the mean decay rate of the reactor-ON period. This period is chosen as the time slot 28~h~--~32~h in Fig.~\ref{fig:Rate_Analysis}. Several peak features are clearly visible in the spectrum shown in Fig.~\ref{fig:energy_spectrum} and they are very similar to the predicted pattern in terms of number of peaks and relative intensities. An \textit{ad hoc} alignment of the predicted and measured peak positions leads to the energy scale curve shown in Fig.~\ref{fig:calib_curve}. The large deviation from linearity is expected from an operating point set quite high in the transition from superconducting to normal phases. Moreover, this calibration curve is confirmed by its very good agreement with an independent calibration point established by the detection of the 59.5~keV $\gamma$-ray line from an intense 3~GBq $^{241}$Am source placed outside the cryostat. This calibration is applied to the data to convert the reconstructed amplitudes from V to MeV. As a final step, the simulated spectrum is convolved by a Gaussian resolution function fitted to the width of the experimental peaks. 

\subsection{Updated decay-scheme of $^{187}$W}
\label{subsec:decay-scheme}

\begin{figure}
    \centering
    \includegraphics[width=\linewidth]{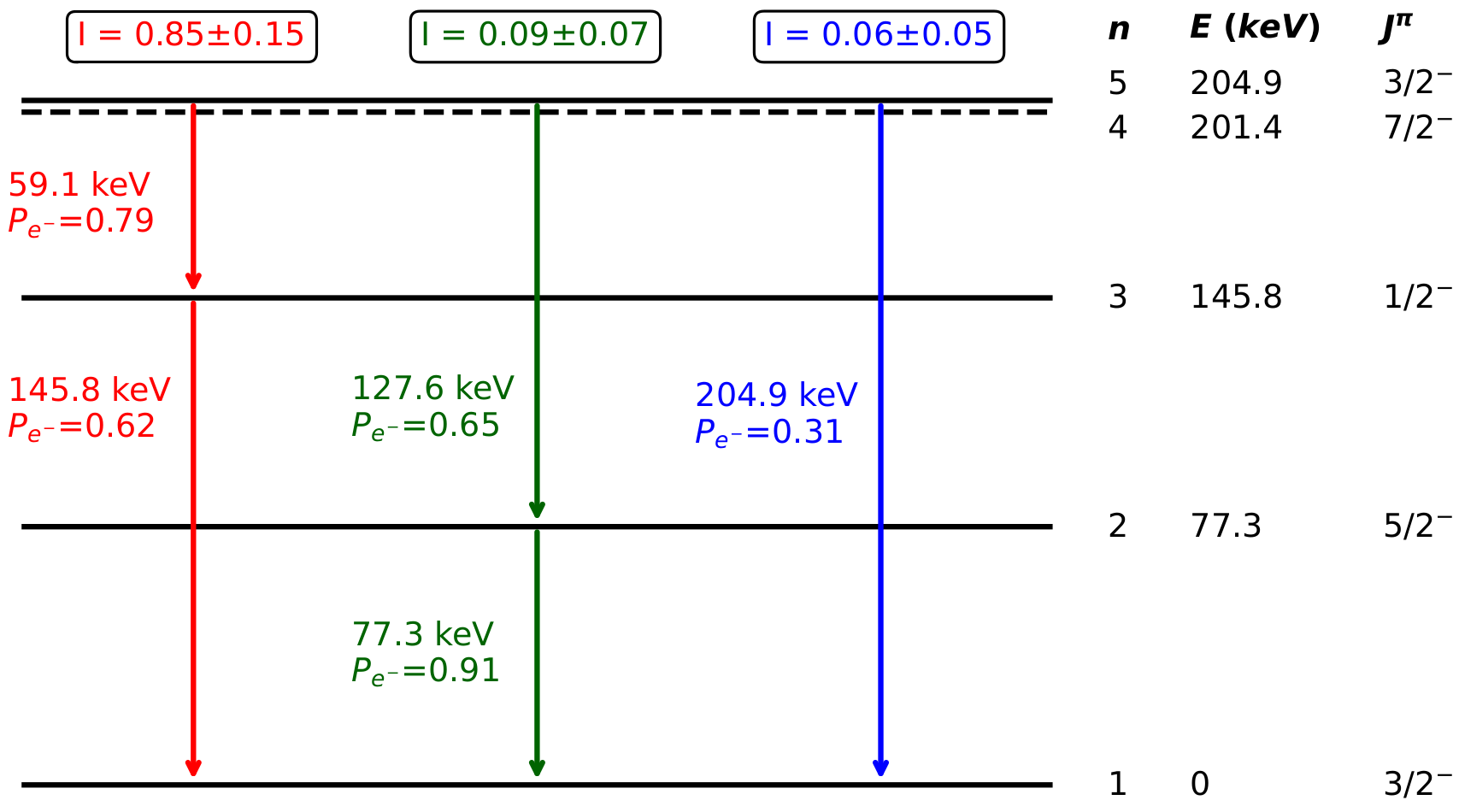}
    \caption{Updated decay scheme of the 5$^\text{th}$ energy level of $^{187}$W deduced from the experimental spectrum in Fig.~\ref{fig:energy_spectrum}. On the right side, in black, the spin, parity and energy of the nuclear levels are indicated. The three possible decay-chains from the 5$^\text{th}$ excited level are highlighted in red, green and blue with their relative intensity $I$ and for each transition the probability $P_{e^{-}}$ to emit a conversion electron instead of a $\gamma$-ray. The intensity of the red decay-chain, set to 0 in the RIPL-3 nuclear database (version 2023) due to the lack of measurement of the 59.1~keV transition, is now found to be the most probable decay.}
    \label{fig:decay_scheme}
\end{figure}

A very good agreement with the data is obtained, validating the refined description of the de-excitation cascades by our simulation. However, in its initial version, the predicted spectrum missed a peak observed in the data at an energy around 60~keV with a statistical significance of 12.2$\sigma$ (see inset in Fig.~\ref{fig:energy_spectrum}). Upon checking the nuclear database RIPL3 (version 2023)~\cite{RIPL3}, used as input in the \fifrelin simulations, we found that the intensity of one of the decay chains from the fifth excited state of $^{187}$W (204.9~keV, 3/2$^{-}$) to the ground state was set to 0 due to the lack of data. This decay chain consists of the emission of two $\gamma$-rays, one of 59.1~keV and the other of 145.8~keV (see Fig.~\ref{fig:decay_scheme}). Thus, the extra peak in the data is interpreted as the fraction of these decays with the 145.8~keV $\gamma$-ray escaping from the cryogenic detector while the energy of the first $\gamma$-ray is fully deposited. Since this fraction depends on the electron conversion coefficients, taken from reference calculations \cite{KIBEDI2008}, and the probability of interaction in the crystal of the considered electrons and $\gamma$-rays, which can be determined by dedicated \geant simulations, an updated decay scheme of $^{187}$W can be inferred with propagated uncertainties. Thus, we propose the following updated transition probabilities from the 5$^{\text{th}}$ excited state of $^{187}$W: $I_{\text{59~keV}}^{5 \rightarrow 3} = 0.85 \pm 0.15$, $I_{\text{127~keV}}^{5 \rightarrow 2} = 0.09 \pm 0.07$ and $I_{\text{205~keV}}^{5 \rightarrow 1} = 0.06 \pm 0.05$ (see Fig.~\ref{fig:decay_scheme}). The relative intensity of the previously omitted decay chain becomes dominant, and the energy spectrum predicted with this new decay scheme is in good agreement with the data. The only other measurement that we could find in literature \cite{PhysRevC.89.014606} suggests a slightly different set of branching ratios ($I_{\text{127~keV}}^{5 \rightarrow 2} = 0.64$, $I_{\text{127~keV}}^{5 \rightarrow 2} = 0.24$ and $I_{\text{205~keV}}^{5 \rightarrow 1} = 0.12$). The energy spectrum simulated with these numbers leads to a 21\% deficit in the intensity of the 59.1~keV peak with respect to the data, with a 1.8$\sigma$ significance. This sensitivity to the input branching ratios is in agreement with the uncertainties quoted in Fig.~\ref{fig:decay_scheme}. 

\subsection{First coincidence data}
\label{subsec:coinc_data}

As the events in the energy spectrum of Fig.~\ref{fig:energy_spectrum} are generated by neutron captures, it is possible for an energetic $\gamma$-ray from the same de-excitation cascade to reach one of the \BaF detectors. These coincidences are very similar to those ultimately sought between a low-energy nuclear recoil and a high-energy $\gamma$-ray. They serve here as a validation of the synchronization between the two data acquisitions and the total detection efficiency of the \BaF detector crown.

For the time being, the synchronization of the cryogenic and $\gamma$-detectors is ensured by an external generator sending a pulse every second to a dedicated channel of each acquisition. A large relative drift between the 10~MHz FASTER clock and the 1~MHz VDAQ3 clock was measured corresponding to a shift of 63.1~$\mu$s/s. We chose as a reference the clock of the cryogenic detector acquisition, with T $=0$ defined by the time of the first external pulse. A time correction is applied in the processing of the $\gamma$-detector data. The result of the processing is saved in a HDF5 file \cite{hdf5} to be used for the coincidence analysis between $\gamma$-detectors and cryogenic detectors. It contains, for all events in the range of interest, the reconstructed $\gamma$-energy, the triggered \BaF counter(s), the raw event time, and the time converted to the reference cryogenic detector time frame. This synchronization method was validated using a second pulse generator plugged in independent cryogenic and $\gamma$-detector channels, and showing that its pulses were still reconstructed in coincidence after several hours of data taking, with a time jitter of 60~$\mu$s. This time precision is sufficient given the $O(100~\mu\text{s})$ rise time of the cryogenic detectors currently tested.

Unlike the events in \CRAB calibration peaks, associated with mono-energetic $\gamma$-rays, the simulation predicts a continuous distribution for the \BaF counterpart of the cryogenic detector spectrum in Fig.~\ref{fig:energy_spectrum}. On the other hand, the $\gamma$-ray spectrum of single events measured in \BaF detectors (Fig. \ref{fig:Gamma_Background}) enables us to determine the spectrum of accidental coincidences. Combining the two information, we define a favorable window of 3.5 to 6~MeV to search for coincidence events induced by neutron captures. 
\begin{figure}[t]
    \centering
    \includegraphics[width=1.0\linewidth]{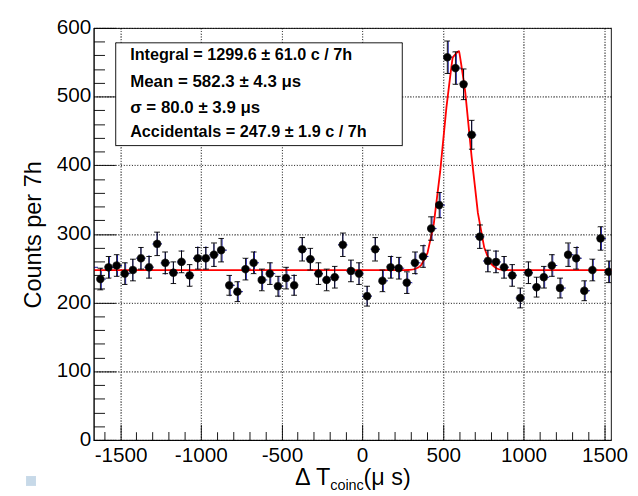}
    \caption{Distribution of the time difference between events in the cryogenic detector W1-682-9 in the 10--300~keV range and events in the \BaF detectors in the 3.5--6.0~MeV range.}
    \label{fig:Coinc_peak}
\end{figure}
Figure~\ref{fig:Coinc_peak} shows the measured distribution of time differences between cryogenic detector events in the 10--300~keV range and \BaF detector events in the 3.5--6.0~MeV range. A coincidence peak is clearly visible above a flat background of accidentals, which level is compatible with the simple expectation of the single rates times the coincidence time window. The coincidence peak is well fitted with a Gaussian function with a standard deviation of 80~$\mu$s, similar to the time jitter observed with the pulse generator. The peak is shifted in time by 582~$\mu$s with respect to the synchronization established with the pulse generator. This delay is attributed to the SQUID readout electronics, through which the pulser signal does not pass when synchronizing both acquisitions. The excess of true coincidences above accidentals has a high statistical significance of more than 20$\sigma$. The ability to detect such coincidences is a major validation of our experimental setup, opening perspectives of high-precision measurements (see Sect. \ref{sec:physics}). 

The measured coincidence rate is (1300\,$\pm$\,61) events per 7~h. After correcting for the aging effect on the \BaF detection efficiency mentioned in Sect. \ref{subsec:BaF2_characteristics} and normalizing to the neutron flux from the rate analysis, the predicted rate is 1455 events for 7~h. This 2.6$\sigma$ discrepancy between the data and the prediction may point to an energy dependence of the aging effect in the \BaF detectors, which will be further investigated by complementary measurements.

\section{Physics program}
\label{sec:physics}

The proofs of concept established with these commissioning data lay a solid foundation for our precision measurement program. In this section, we present an update of the expected rates and sensitivities, taking into account the actual experimental configuration installed on site and the measured background levels.

\subsection{Candidate cryogenic detectors}
\label{subsec:fom}

An online repository of the output of \fifradina simulations is regularly updated \cite{soum_sidikov_2023_7936552}. It currently contains predictions for four detector materials (\CaWO, \AlO, Ge and Si), covering most detector technologies in use or in development in the community. Other target nuclides may be studied in the future. An online version of \fifrelin is available for preliminary studies~\cite{FIFRELIN_userGuide}, together with the library for interfacing output files with \geant~\cite{thulliez_loic_2023_7933381}.

The main calibration lines expected from neutron capture in the above-mentioned materials have already been presented in \cite{Soum2023} and are listed in greater details in Table~\ref{tab:calibration_lines}. As seen in this table, deexcitation cascades with two or three $\gamma$-rays can actually generate recoil lines when the lifetime of the intermediate nuclear levels is longer than the nucleus stopping time in matter (Eq.~\ref{eq:slow_hypothesis}). In order to compare the event rate in each recoil line we introduce an improved Figure of Merit ($FoM$) with respect to the quantity defined in \cite{Soum2023}
\begin{equation}
    FoM(\text{X},\gamma_\text{c}) = \Sigma_{(\text{n},\gamma)}(\text{X}) \times I_{\gamma_\text{c}}
\end{equation}
with $\Sigma_{(\text{n},\gamma)}(\text{X})$ the macroscopic cross section of radiative neutron capture for a target isotope $\text{X}$ and $I_{\gamma_\text{c}}$ the branching ratio to the $\gamma_\text{c}$ de-excitation cascade of the compound nucleus after neutron capture. $\Sigma_{(\text{n},\gamma)}(\text{X})$ is the product of the microscopic cross section multiplied by the number density of the target isotope $\text{X}$. It can be interpreted as the probability per unit path length that a neutron will be captured by this nuclide. 
Note that $FoM$ is not directly proportional to the count rate expected for each recoil line, as we still need the condition that all particles from the $\gamma_\text{c}$ de-excitation cascade escape the crystal without interacting. The validity of this condition depends on the geometry of each detector and will therefore have to be addressed specifically for each concrete case. Therefore, the last column of Table~ \ref{tab:calibration_lines} shows the expected rates for a given geometry of each material, chosen as a representative example of existing detectors. In the case of silicon, the detector considered is larger to compensate for the lower neutron capture cross section. This will affect the energy resolution but the higher energy of the recoil peaks, in the keV range, provides some margin. Based on the energy resolution already achieved in existing detectors, all predicted rates allow in principle an accurate measurement of the mean position within a reasonable run time considering the 35 hours per week of reactor operation of the Vienna TRIGA Mark-II reactor. In the following, when two lines become too close to be detected separately, the average position will be the sum of the individual positions weighted by the count rates. 

\bgroup
\renewcommand*{\arraystretch}{1.5}
\begin{table*}
\caption{Summary of the most prominent nuclear recoil peaks induced by radiative thermal neutron capture in four detector materials of various sizes (1st column). The 2nd column shows the macroscopic capture cross section $\Sigma_{(\text{n},\gamma)}$, sum of $\Sigma_{(\text{n},\gamma)}(\text{X})$ over all isotopes $\text{X}$ in the material. The target isotopes are listed in the 3rd column with their relative probability $P_{\text{capture}}$ of neutron capture (4th column), defined as the ratio $\Sigma_{(\text{n},\gamma)}(\text{X})$/$\Sigma_{(\text{n},\gamma)}$. The de-excitation of the compound nucleus is characterized by the branching ratio $I_\gamma$ (5th column) and energy $E_\gamma$ (6th column) for each transition, extracted from the RIPL-3~\cite{RIPL3} and EGAF~\cite{EGAFpubli} databases. Multi-$\gamma$ cascades span multiple rows, with each row describing one $\gamma$-ray of the cascade. Subsequent emissions are signaled by a $\hookrightarrow$ symbol preceding their energy, and the 7th column indicates the associated half-life. The reported half-lives are evaluated values from RIPL-3 database when available,
with the Weisskopf estimate used as a fallback, and specified by the notation "(W)". The three last columns present the total recoil energy $E_\text{r}$, the Figure of Merit $FoM$ and the expected counting rate associated to a given cascade, assuming the detector geometry shown in the 1st column, 140 hours of data taking with 100\% efficiency and full beam intensity of 442~cm$^{-2}$.s$^{-1}$, except for the Ge detector where a reduced intensity of 105~cm$^{-2}$.s$^{-1}$ is used to keep the total rate in the detector at 2~Hz. Table taken from \cite{Soum_PhD_thesis}.}
\label{tab:calibration_lines}
\begin{center}
\begin{tabular*}{\textwidth}{@{\extracolsep{\fill}}cc|cc|cccccc@{}}
\hline
  \multicolumn{2}{c|}{\textbf{Detector material}} & \multicolumn{2}{c|}{\textbf{Target nucleus}} & \multicolumn{6}{c}{\textbf{Compound nucleus}} \\

  \multirow{2}{*}{\shortstack[1]{Formula \&\\ Size (mm)}} &   $\Sigma_{(\text{n}, \gamma)}$ &\multirow{2}{*}{Nuclide}  & $P_\text{capture}$ & $I_\gamma$& $E_\gamma$ & Half-life & $E_\text{r}$ & $FoM$ & Rate\\
     & $(\text{cm}^{-1})$ & & $(\%)$ & $(\%)$ & $($MeV$)$ & $($ps$)$ & $($eV$)$ & ($\times 10^{4}$) & (mHz)   \\
\hline\hline
\multirow{17}{*}{\shortstack[1]{\AlO\\$5\times5\times5$}} & \multirow{17}{*}{$1.10\times10^{-2}$} & \multirow{2}{*}{${}^{27}$Al} & \multirow{2}{*}{99.89} & 6.90 & 4.133 & - & \multirow{2}{*}{571.0} & \multirow{2}{*}{0.89} & \multirow{2}{*}{0.4} \\
  &  &  &  & 11.73 & $\hookrightarrow$ 3.560 & 0.03 &  &  &  \\
  \cline{3-10}
 &  & \multirow{2}{*}{${}^{27}$Al} & \multirow{2}{*}{99.89} & 0.56 & 3.825 & - & \multirow{2}{*}{572.0} & \multirow{2}{*}{0.28} & \\
  &  &  &  &46.08 & $\hookrightarrow$ 3.902 & 0.19 &  &  &  \\
  \cline{3-9}
 &  & \multirow{2}{*}{${}^{27}$Al} & \multirow{2}{*}{99.89} & 3.12 & 3.849 & - & \multirow{2}{*}{572.0} & \multirow{2}{*}{2.68} &  \\
  &  &  &  &78.06 & $\hookrightarrow$ 3.876 & 0.02 &  &  &  \\
    \cline{3-9}
 &  & \multirow{2}{*}{${}^{27}$Al} & \multirow{2}{*}{99.89} & 0.87 & 3.789 & - & \multirow{2}{*}{572.0} & \multirow{2}{*}{0.39} & \multirow{-4}{*}{0.4} \\
  &  &  &  &40.65 & $\hookrightarrow$ 3.936 & 0.02 &  &  &  \\
  \cline{3-10}
 &  & \multirow{2}{*}{${}^{27}$Al} & \multirow{2}{*}{99.89} & 0.73 & 4.015 & - & \multirow{2}{*}{573.0} & \multirow{2}{*}{0.30} & \multirow{2}{*}{0.2} \\
  &  &  &  &37.17 & $\hookrightarrow$ 3.705 & 0.19 &  &  &  \\
\cline{3-10}
 &  & \multirow{2}{*}{${}^{27}$Al} & \multirow{2}{*}{99.89} & 6.90 & 4.134 & - & \multirow{2}{*}{575.0} & \multirow{2}{*}{4.49} & \multirow{2}{*}{0.8} \\
  &  &  &  & 59.27 & $\hookrightarrow$ 3.902 & 0.03 &  &  &  \\
\cline{3-10}
 &  & \multirow{2}{*}{${}^{27}$Al} & \multirow{2}{*}{99.89} & 6.80 & 4.260 & - & \multirow{2}{*}{578.0} & \multirow{2}{*}{6.55} & \multirow{2}{*}{0.9} \\
  &  &  &  & 87.72 & $\hookrightarrow$ 3.466 & 0.04 &  &  &  \\
\cline{3-10}
  &  & \multirow{2}{*}{${}^{27}$Al} & \multirow{2}{*}{99.89} & 3.39 &7.693 & - & \multirow{2}{*}{1135.7} & \multirow{2}{*}{3.72} & \multirow{2}{*}{2.0} \\
  &  &  &  & 100 & $\hookrightarrow$ 0.031 & 2070 &  &  &  \\
\cline{3-10}
  &  & ${}^{27}$Al & 99.89 & 26.81 & 7.724 & - & 1144.8 & 29.46 & 30.5 \\
\hline \hline
\multirow{4}{*}{\shortstack[1]{Si\\$10\times10\times20$}} & \multirow{4}{*}{$8.23\times 10^{-3}$} & \multirow{2}{*}{${}^{28}$Si} & \multirow{2}{*}{94.60} & 7.10 & 7.200 & - & \multirow{2}{*}{990.4} & \multirow{2}{*}{5.53} & \multirow{2}{*}{46.3} \\
 &  &  &  & 100 & $\hookrightarrow$1.273 & 0.29 & &  & \\
\cline{3-10}
 &  & ${}^{28}$Si & 94.60 & 2.17 & 8.474 & - & 1330.1 & 1.69 & 26.7 \\
\cline{3-10}
 & & ${}^{29}$Si & 3.38 & 6.73 & 10.609 & - & 2016.0 & 0.19 & 3.1 \\
\hline

\end{tabular*}
\end{center}
\end{table*}

\begin{table*}
    \begin{center}
        \begin{tabular*}{\textwidth}{@{\extracolsep{\fill}}cc|cc|cccccc@{}}
\hline
  \multicolumn{2}{c|}{\textbf{Detector material}} & \multicolumn{2}{c|}{\textbf{Target nucleus}} & \multicolumn{6}{c}{\textbf{Compound nucleus}} \\

  \multirow{2}{*}{\shortstack[1]{Formula \&\\ Size (mm)}} &   $\Sigma_{(\text{n}, \gamma)}$ &\multirow{2}{*}{Isotope}  & $P_\text{capture}$ & $I_\gamma$& $E_\gamma$ & Half-life & $E_\text{r}$ & $FoM$ & Rate\\
     & $(\text{cm}^{-1})$ & & $(\%)$ & $(\%)$ & $($MeV$)$ & $($ps$)$ & $($eV$)$ & ($\times 10^{4}$) & (mHz)   \\
\hline \hline

\multirow{15}{*}{\shortstack[1]{Ge\\$10\times10\times10$}} & \multirow{15}{*}{$9.76\times 10^{-2}$} & \multirow{2}{*}{${}^{74}$Ge} & \multirow{2}{*}{8.58} & 11.75 & 6.253 & - & \multirow{2}{*}{280.6} & \multirow{2}{*}{9.69} & \multirow{2}{*}{12.5} \\
 &  &  &  & 98.52 & $\hookrightarrow$0.253 & 1.36 (W) & &  & \\
\cline{3-10}
 &  & \multirow{2}{*}{${}^{70}$Ge} & \multirow{2}{*}{28.30} & 5.30 & 6.117 & - & \multirow{2}{*}{296.0} & \multirow{2}{*}{11.51} & \multirow{2}{*}{8.8} \\
 &  &  &  & 78.65 & $\hookrightarrow$1.299 & 0.4 & &  & \\
 \cline{3-10}
 &  & ${}^{74}$Ge &  8.58 & 2.83 & 6.506 & - & 303.2 & 2.37 & 4.5 \\
 \cline{3-10}
 &  & \multirow{2}{*}{${}^{70}$Ge} & \multirow{2}{*}{28.30} & 2.62 & 6.276 & - & \multirow{2}{*}{307.9} & \multirow{2}{*}{5.59} & \multirow{2}{*}{6.8} \\
  &  &  &  & 77.31 & $\hookrightarrow$1.139 & 4.00 &  &  & \\
 \cline{3-10}
 &  & \multirow{2}{*}{${}^{70}$Ge} & \multirow{2}{*}{28.30} & 4.80 & 6.708 & - & \multirow{2}{*}{344.3} & \multirow{2}{*}{12.65} & \multirow{2}{*}{18.2} \\
 &  &  &  & 95.44 & $\hookrightarrow$0.708 & $<$10.70 &  &  & \\
\cline{3-10}
&  & \multirow{2}{*}{${}^{70}$Ge} & \multirow{2}{*}{28.30} & 3.83 & 6.916 & - & \multirow{2}{*}{363.9} & \multirow{2}{*}{10.50} & \multirow{2}{*}{13.7} \\
 &  &  &  & 99.30 & $\hookrightarrow$0.500 & 0.18 (W) &  &  & \\
\cline{3-10}
 &  & ${}^{70}$Ge & 28.30 & 1.95 & 7.416 & - & 416.2 & 5.39 & 9.5 \\
\cline{3-10}
 &  & \multirow{3}{*}{${}^{73}$Ge} & \multirow{3}{*}{51.55} & 1.02 & 8.732 & - & \multirow{3}{*}{561.8} & \multirow{3}{*}{5.12} & \multirow{3}{*}{5.1} \\
 &  &  &  & 99.95 & $\hookrightarrow$0.868 & 1.53 &  &  & \\
 &  &  &  & 99.86 & $\hookrightarrow$0.596 & 12.14 &  &  & \\
\hline \hline

\multirow{9}{*}{\shortstack[1]{\CaWO{}\\$4.8\times4.8\times4.8$}} & \multirow{9}{*}{$2.37\times 10^{-1}$} & \multirow{2}{*}{${}^{186}$W} & \multirow{2}{*}{58.11} & 7.42 & 5.262 & - & \multirow{2}{*}{79.6} & \multirow{2}{*}{28.34} & \multirow{2}{*}{13.4} \\
 &  &  &   & 27.73 & $\hookrightarrow$0.205 & 2.6 (W) &  &  &  \\
\cline{3-10}

  &  & \multirow{2}{*}{${}^{182}$W} & \multirow{2}{*}{29.01} & 5.24 & 5.165 & - & \multirow{2}{*}{81.3} & \multirow{2}{*}{18.84} & \multirow{2}{*}{13.4} \\

 &  &  &   & 52.3 & $\hookrightarrow$1.026 & - &  &  &  \\

\cline{3-10}

 &  & \multirow{2}{*}{${}^{186}$W} & \multirow{2}{*}{58.11} & 5.26 & 5.321 & - & \multirow{2}{*}{81.4} & \multirow{2}{*}{27.68} & \multirow{2}{*}{7.8} \\

 &  &  &   & 38.21 & $\hookrightarrow$0.146 & 7.1 (W) &  &  &  \\
 
\cline{3-10}
 & & ${}^{186}$W & 58.11 & 0.26 & 5.467 & - & 85.8 & 3.58 & 3.4 \\
\cline{3-10}
 &  & ${}^{182}$W & 29.01 & 13.94 & 6.191 & - & 112.5 & 95.84 & 84.3 \\
\cline{3-10}
 & & ${}^{183}$W & 7.62 & 5.83 & 7.411 & - & 160.3 & 10.53 & 9.8 \\
\hline   
        \end{tabular*}
    \end{center}
\end{table*}

\egroup

Our short-term objective is to measure nuclear recoil peaks in \AlO and \CaWO using the \NUCLEUS detectors already available. The expected statistical accuracy is better than 1\%, and the nuclear energy scale thus obtained can be compared with the energy scale of the electron recoils induced either by the recently installed LED light pulse injection device \cite{DELCASTELLO2024169728}, or by a fluorescence X-ray device such as the one developed by \NUCLEUS. Then a program of measurements on different germanium detector technologies is planned, see Sect. \ref{subsec:quenching}. In the following, we discuss also some recent updates to our simulations combined with the commissioning benchmark measurements enabling us to envisage an original program to gain a detailed understanding of the response of cryogenic detectors, with applications to solid-state physics.

\subsection{Refined tests of timing effects in \AlO}
\label{subsec:Al2O3_case}
\begin{figure}[t]
    \centering
    \includegraphics[width=1.0\linewidth]{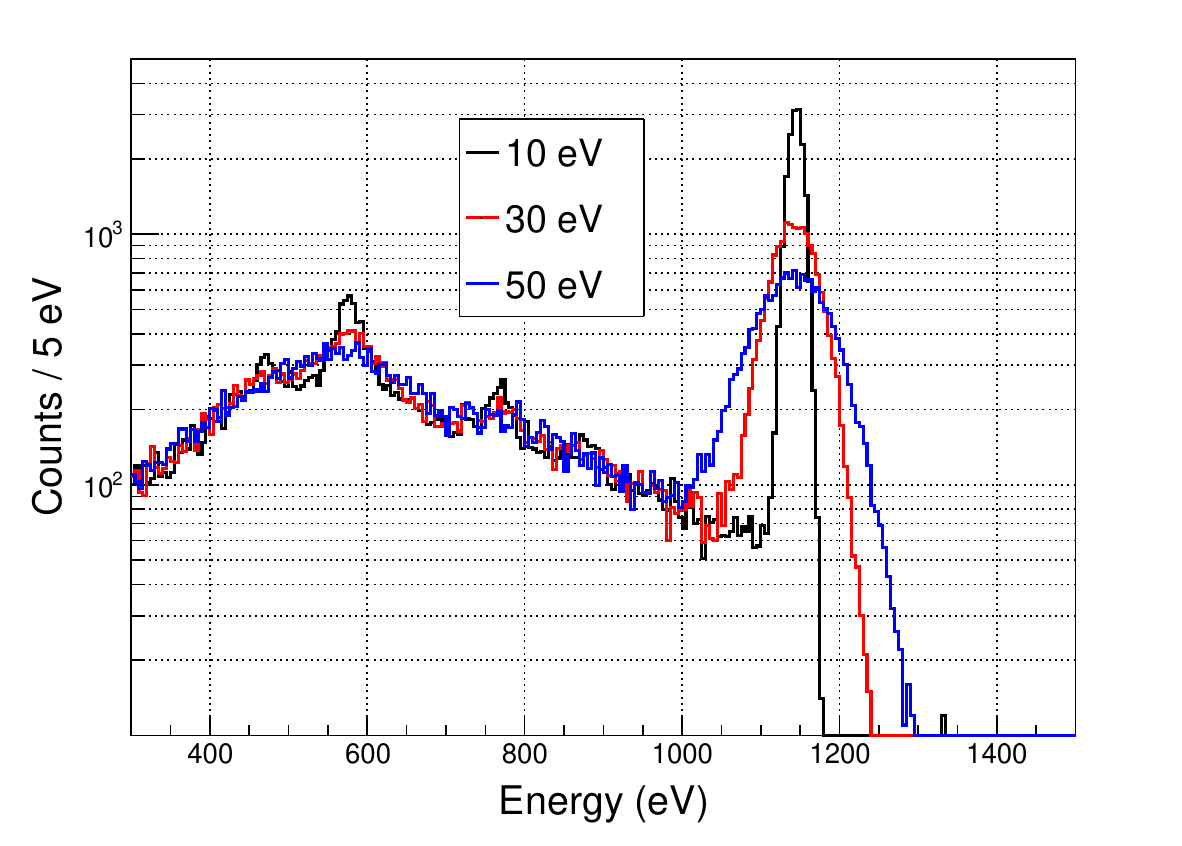}
    \caption{Updated \fifradina simulations of nuclear recoil spectra induced by neutron capture in a \AlO cryogenic detector ($5\times5\times5$ mm$^{3}$) with one month of data taking at full efficiency (equivalent to 140~h) and for different energy resolutions. In each case a constant resolution is applied across the whole energy range.}
    \label{fig:Al2O3_spectra}
\end{figure}

The review of all simulations with the experimental configuration at ATI in Vienna has led to the correction of two inaccuracies in the prediction of the \AlO recoil spectrum with respect to what was presented in \cite{Soum2023}: the atom density input value of \iradina simulations has been updated and partial lifetimes (nuclear level lifetime divided by the transition probability) have been used for each $\gamma$-transition instead of the total lifetime of the parent nuclear level. This has little impact on most of the recoil spectrum, with the exception of a new peak emerging at 575~eV (see Fig.~\ref{fig:Al2O3_spectra}). This energy is actually a weighted average of the first seven recoil lines reported in Table \ref{tab:calibration_lines}. Similarly to the case of silicon discussed in \cite{Soum2023}, this peak results from comparable values of the second $\gamma$-transition lifetime and the aluminum stopping time in \AlO. 
According to Table~\ref{tab:calibration_lines} it should be detectable with high significance ($>$10$\sigma$) in a \NUCLEUS \AlO detector within a month of data taking (140 h of reactor operation) and up to $\approx$ 30~eV resolution. 

Sub-percent statistical precision is achievable for both peaks at 575 and 1145~eV, providing an accurate calibration of the response to nuclear recoils in the sub-keV range. 
Beyond this calibration the relative amplitude between the 575~eV peaks and the single-$\gamma$ peak at 1145~eV will be very sensitive to the timing effects, thus testing our understanding about the interplay between lifetime of nuclear levels and of stopping times in matter.

\subsection{Sensitivity to the formation of crystal defects in \CaWO}
\label{subsec:defects}

After the \fifradina and \geant simulations, a final step is in principle still required before predicted and measured spectra can be compared: in the cascade of atomic displacements generated by the PKA, part of the initial kinetic energy can be stored in crystal defects instead of being fully converted to the usual readout channels of the detectors (heat, ionization, and scintillation). A fraction of the initial energy thus becomes invisible to detectors.

Above $\approx$1~keV, the number of defects created is sufficiently large and the main effect is a missing energy, which depends linearly on the initial energy. This results in a simple renormalization of the calibration coefficient, with little expected spectral distortion. Lower energies become sensitive to threshold effects. First, there is a threshold displacement energy, below which no defect at all is created. Then comes a transient regime that gradually enable the various types of defect (\textit{e.g.} local lattice distortions, broken atomic bonds, vacancy-interstitial pairs). In this transient zone, the response of the cryogenic detector will be non-linear, implying distortions in the measured spectra \cite{PhysRevD.106.063012}. By definition, the position of the \CRAB nuclear recoil peaks includes all these effects, whereas they will be absent from calibrations based on electronic recoils.

\begin{figure}[h]
    \centering
    \includegraphics[width=1.0\linewidth]{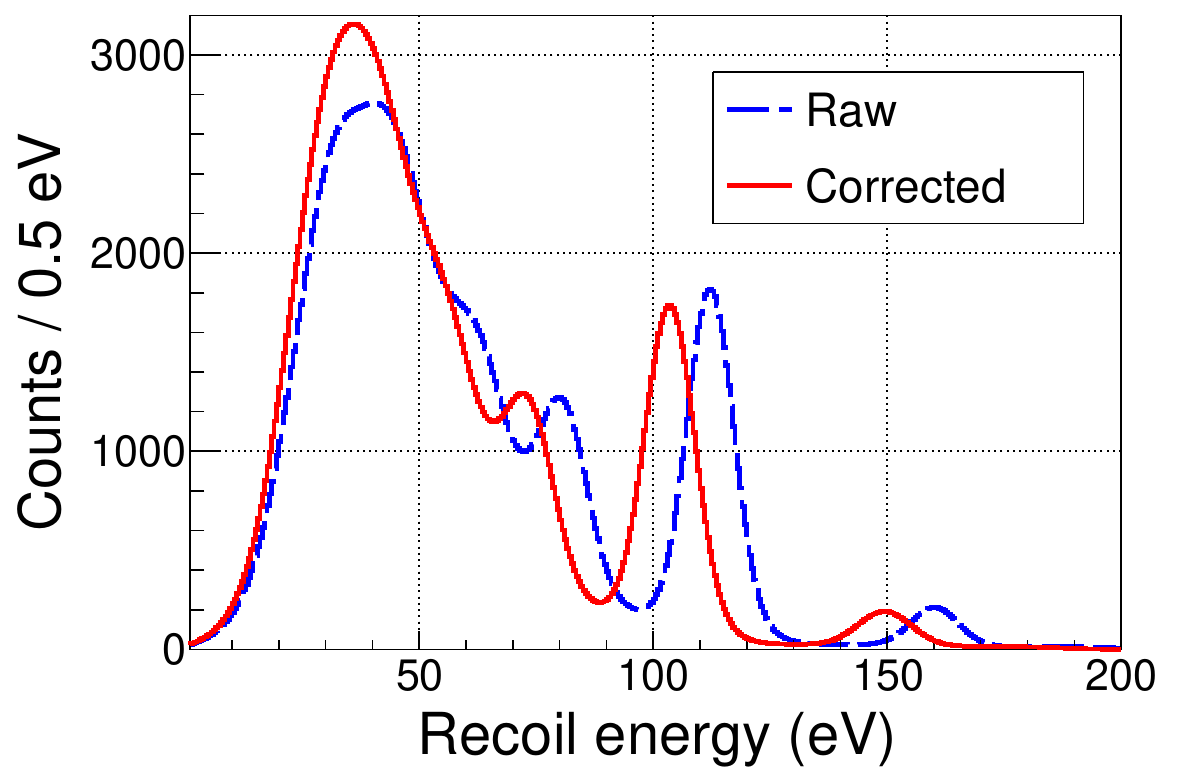}
    \caption{Predicted nuclear recoil spectra induced by neutron capture in a \CaWO cryogenic detector ($4.8\times4.8\times4.8$~mm$^{3}$) for one month of data taking at full efficiency (equivalent to 140~h) with 5eV energy resolution (dashed blue curve). 
    The effect of the energy stored in crystal defects computed in MD simulations is implemented following the procedure described in \cite{soumsidikov2024} (solid red curve).}
    \label{fig:CaWO4_Defects}
\end{figure}

Molecular dynamics (MD) calculations of tungsten recoils in \CaWO have recently been carried out by constructing a Machine Learning interatomic potential trained on a database of defect configurations computed with the Density Functional Theory \cite{soumsidikov2024}. They establish the displacement threshold in \CaWO at 40~eV. Therefore, the creation of defects should induce measurable spectral distortions in the energy range probed by the \CRAB spectrum (E $<$ 160~eV) as illustrated in Fig.~\ref{fig:CaWO4_Defects}. The shift between spectra with and without crystal defects is of course not experimentally accessible, but the distortion in shape could be tested, offering an original probe of the fundamental processes of defect creation.

\begin{figure}[t]
    \centering
    \includegraphics[width=1.0\linewidth]{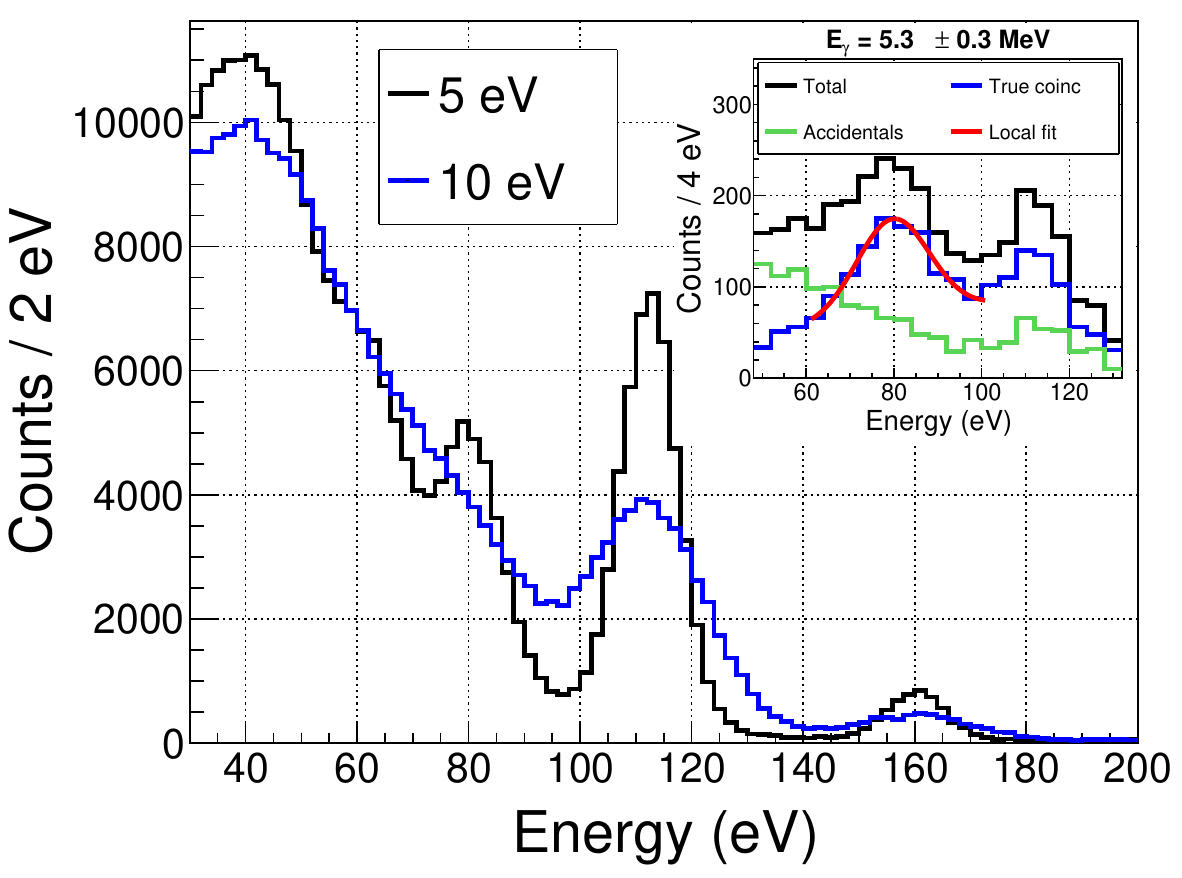}
    \caption{Predicted nuclear recoil spectra induced by neutron capture in a \CaWO cryogenic detector ($4.8\times4.8\times4.8$~mm$^{3}$) for one month of data taking at full efficiency ( equivalent to 140 h). The 80~eV recoil peak visible with 5~eV energy resolution (black curve) is smeared out with 10~eV resolution. Inset: requiring a coincidence with an energy deposition of $E_\gamma = (5.30\pm 0.28)$~MeV in a $\gamma$-detector allows to recover this peak structure for the same resolution of 10~eV. The spectral shape of accidental coincidences (green curve) is accurately measured by the single spectrum in the $\gamma$-detectors and can be subtracted (blue curve) from the total spectrum (black curve) without adding extra statistical uncertainty. The peak at 112.5~eV, associated with higher energy $\gamma$-rays, is also visible due to Compton scattering of these $\gamma$-rays in the \BaF crystals.}
    \label{fig:CaWO4_coinc_spectrum}
\end{figure}

Another approach is to use a basic local fit (Gaussian + 1st order polynomial function) to measure the relative positions of the three main peaks at 80.8, 112.5 and 160.3~eV and demonstrate the deviation from linearity induced by the defects since the predicted mean stored energies for each of these recoils are 7.5, 8.9 and 11.0~eV. However, a meaningful measurement of this effect requires a very good energy resolution, since the 80~eV peak is no longer visible at 10~eV resolution (Fig.~\ref{fig:CaWO4_coinc_spectrum}). Statistical analysis shows that with 140~h of reactor data, equivalent to one month of data taking at full efficiency, a resolution of at least 6.5~eV is required to maintain significance greater than 5$\sigma$ \cite{Soum_PhD_thesis}. The detection in coincidence of the primary $\gamma$-rays associated with the nuclear recoils relaxes this constraint on energy resolution. The case illustrated in the inset of Fig.~\ref{fig:CaWO4_coinc_spectrum} shows that with 10~eV energy resolution the 80~eV peak can be measured with $\approx$ 7$\sigma$ significance in 140~h.

\subsection{Study of nuclear ionization yield in Ge and Si}
\label{subsec:quenching}
The 1 cm$^3$ germanium detector listed in Table \ref{tab:calibration_lines} is representative of the technology of low threshold detector being developed by the \TESSERACT collaboration~\cite{Billard:2024zvc}. Inherited from the \EDELWEISS~\cite{EDELWEISS:2017lvq} and \RICOCHET~\cite{Ricochet:2022pzj} experiments, it offers simultaneous reading of signals in the ionization and phonon channels. Indeed, following a particle's interaction in the detector medium, the induced recoil will release its energy by creating both phonons (heat) and charge carriers (ionization). To first order\footnote{We neglect here the phonon energy loss due to Frenkel defects and to charge trapping, which will also be addressed in the proposed physics program.}, the different measurable energy quantities are intertwined as follows:
\begin{align}
    & E_{\rm ion} = Q(E_\text{r})\, E_\text{r} \ ,\nonumber \\
    & E_{\rm NTL} = E_{\rm ion}\,\frac{V}{\epsilon} \ , \nonumber \\
    & E_{\rm ph} = E_\text{r} + E_{\rm NTL} = E_\text{r}\left[1+Q(E_\text{r})\frac{V}{\epsilon}\right] \ , \nonumber
    \label{eq:GeNTL}
\end{align}
where $E_{\rm ion}$ stands for the ionization energy, related to recoil energy $E_\text{r}$ by the quenching factor $Q(E_\text{r})$. $E_{\rm NTL}$ is the additional Neganov-Trofimov-Luke phonon energy produced by drifting the charge carriers across the crystal~\cite{Luke:1988xcw,Neganov:1985khw}, with $V$ the bias voltage and $\epsilon$ is the average energy required for an electron recoil to produce an electron-hole pair (3.0~eV and 3.8~eV respectively for Ge~\cite{EDELWEISS:2006bcu,knoll2010radiation} and Si~\cite{PEHL196845}). $E_{\rm ph}$ is the phonon energy. Interestingly, semiconducting detectors can be either operated at low voltage (typically $\leq~4$~V), such that $E_{\rm NTL} < E_\text{r}$  for nuclear recoils, or at high voltage ($\sim 100$~V), where the cryogenic calorimeter is turned into a charge amplifier with $E_{\rm ph} \simeq E_{\rm ion}$. Following the recent results from the \RICOCHET collaboration, in the context of a low-voltage operation, ionization and phonon baseline resolutions of 30~eV$_\mathrm{ee}$~\cite{Ricochet30eV} (eV electron-equivalent) and 30~eV~\cite{Ricochet:2021kqt}, respectively, have been achieved with 42~g Ge crystals. These results suggest that 20~eV$_\mathrm{ee}$ and 20~eV ionization and phonon resolutions would be achievable with the proposed 1 cm$^3$ Ge detector prototype to be installed in the \CRAB setup by 2026 with the \RICOCHET readout technology~\cite{Ricochet:2021cqv}. In the context of high-voltage operations, the French \TESSERACT team has recently achieved the first demonstration of a single-electron sensitivity on a 42 g Ge cryogenic detector, suggesting that similar single charge sensitivity would be achievable on a 1 cm$^3$ prototype dedicated to the \CRAB experiment. The combination of both low- and high-voltage approaches, with performance as described above, will offer unprecedented precision measurement of the sub-keV low-energy nuclear recoil energy-scale and ionization yield provided by the \CRAB setup. It is worth noting that similar performance are expected to be achieved with 1 cm$^3$ silicon detectors (see for instance \cite{PhysRevLett.121.051301,Romani:2017iwi,Hong:2019zlm}), hence opening the possibility to extend our study to Si target material.

\begin{figure}[t]
    \centering
    \includegraphics[width=1.0\linewidth]{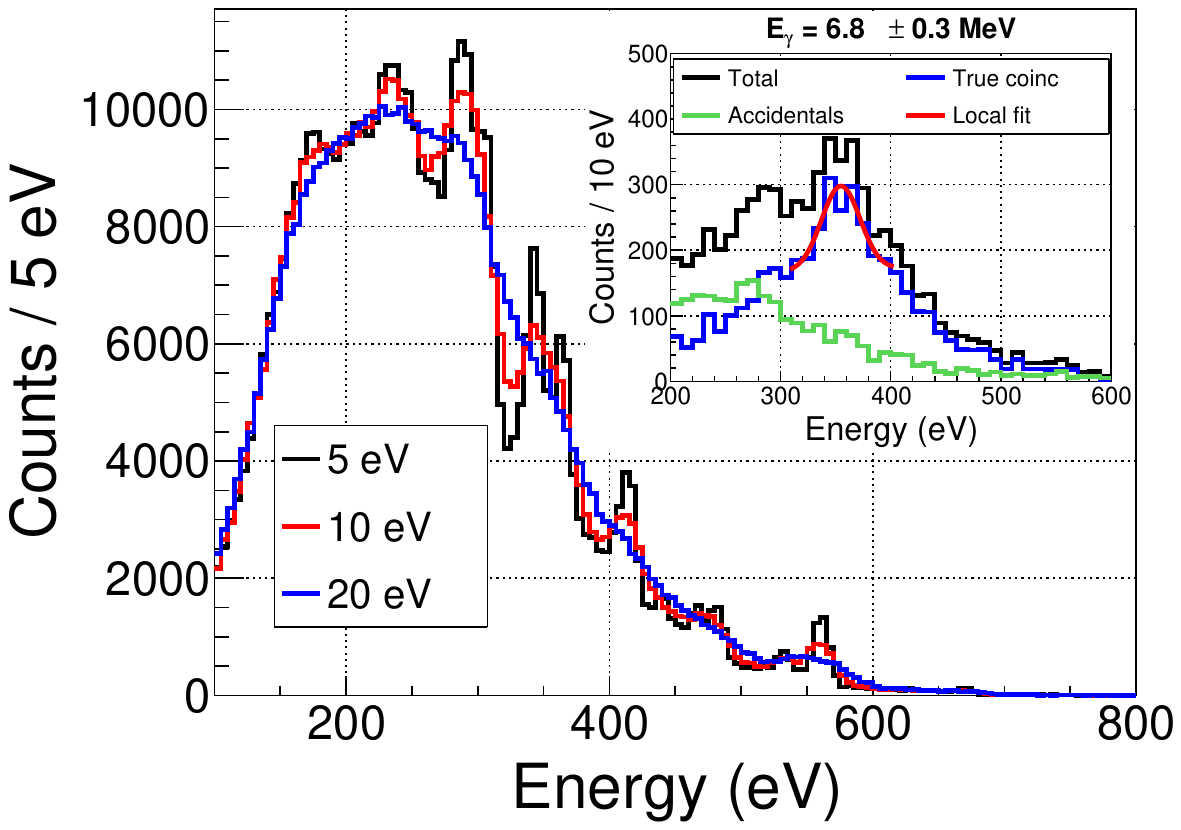}
    \caption{Predicted nuclear recoil spectra induced by neutron capture in a Ge cryogenic detector (1~cm$^{3}$) for one month of data taking at full efficiency (equivalent to 140~h). The neutron beam intensity is reduced to 105~cm$^{-2}$.s$^{-1}$ to keep the total rate in the detector around 2~s$^{-1}$. All recoil lines are smeared out with a 20~eV energy resolution. Inset: selection of events in coincidence with a (6.8$\,\pm\,$0.3)~MeV $\gamma$ associated to the 344 and 364~eV recoil lines. A peak feature is clearly recovered despite the 20~eV resolution.}
    \label{fig:Ge_spectra}
\end{figure}

The predicted spectra of nuclear recoils in a 1~cm$^3$ Ge detector installed in the \CRAB setup for 140~h are shown in Fig.~\ref{fig:Ge_spectra}, for three energy resolutions. Although all recoil lines cannot be resolved, clear peaked structures appear that can be attributed to the triplet of lines at 281, 296 and 303~eV, the doublet at 344 and 364~eV and the two single lines at 416 and 561~eV. However, the exploitation of calibration peaks is conditional on a challenging energy resolution, all spectrum structures being smeared out when applying a 20~eV resolution to the predicted spectrum. Coincident $\gamma$-detection is crucial here to relax this constraint. Indeed, the inset in Fig.~\ref{fig:Ge_spectra} shows the spectrum obtained for the same data set simulated with 20~eV resolution, but imposing the conditions of a (6.8$\pm$0.3)~MeV $\gamma$ event in one of the \BaF detectors. A large fraction of the continuum background is rejected, making it possible to recover the structure of the doublet of lines at 344 and 364~eV, associated with $\gamma$-emission of 6.7 and 6.9~MeV respectively. This is of course at the expense of a severe loss in statistics due to the $\approx$3.2\% total efficiency of the \BaF crowns at 7~MeV. Still, the significance of this calibration peak is estimated at 7$\sigma$ with 140 hours of reactor data.

The simultaneous readout of the ionization and phonon signals for these nuclear recoils of known energy would shed unique light on the study of the ionization yield in germanium but also in silicon, a central topic for future and ongoing experiments. In fact, recent results for the Ge case show widely different and incompatible ionization yield measurements for sub-keV nuclear recoils~\cite{Bonhomme:2022lcz,Collar:2021fcl,Kavner:2024xxd}. It is worth noticing that so far all these sub-keV measurements have been performed using ionization-only HPGe detectors pushing their measurements at the very edge of their noise walls, where accurate signal amplitude reconstructions can be challenging. Interestingly, only the method employed by \cite{Collar:2021fcl} and \cite{Kavner:2024xxd}, and introduced by Jones and Kraner \cite{Jones:1975zze}, measuring the summed ionization energy from a 68.75~keV $\gamma$ and a 254~eV nuclear recoil emitted after a neutron capture in \textsuperscript{73}Ge, is significantly distant from the noise wall, but relies heavily on the systematics of nuclear physics (see \cite{Kavner:2024xxd} for a detailed discussion). Both the low-and high-voltage approaches proposed here will be uniquely positioned to perform such measurements with reduced systematics thanks to dual phonon-ionization readout, with $\sim$100 eV phonon energy threshold, and  single-charge sensitivity, respectively. 
Furthermore, the continuous part of the spectrum itself is a broad peak distribution of pure nuclear recoils. Combined with the nuclear recoil lines seen in Fig.~\ref{fig:Ge_spectra} it will allow extending the ionization yield measurement over the entire recoil energy range from 100~eV to 700~eV.

Similar tensions exist in recent Si ionization measurements~\cite{SuperCDMS:2023geu,Chavarria:2016xsi,Villano:2021eof} in the sub-keV recoil energy range. As for the Ge case, combining our low- and high-voltage approaches, we plan to perform a high-precision measurement of the ionization yield down to 100~eV thanks to the continuous nuclear recoil spectra combined with the nuclear recoil energy lines listed in Table~\ref{tab:calibration_lines}. The proposed ionization yield measurements will have considerable implications to all low-threshold CE$\nu$NS~\cite{Abdullah:2022zue} and light dark matter search~\cite{Billard:2021uyg} experiments using Ge and Si target materials.

\section{Conclusion}
\label{sec:conclusion}
The experimental setup of the \CRAB project installed at the Atominstitut's TRIGA Mark-II reactor in Vienna is an innovative coupling between a collimated thermal neutron beam, a dilution cryostat and \BaF $\gamma$-detectors. The commissioning data presented in this article fully validate the concept of using neutron capture in cryogenic detectors to precisely study their response in the sub-keV energy range. The low-intensity neutron beam meets the specifications with an intensity of $\approx (469 \pm 47)$~cm$^{-2}$s$^{-1}$ and small angular divergence. The Kelvinox 100 dilution refrigerator has been operating stable at nominal temperature of 10~mK for weeks. The first data from a 0.67~g \CaWO detector of the \NUCLEUS collaboration have demonstrated the accurate understanding of the counting rates, including the contributions of neutron captures in the detector crystal and the holder mechanics, the activation-deactivation processes linked to the reactor activity and the external backgrounds. A \%-level agreement between predicted and measured rates was achieved with 3 free parameters only for reactor On and Off periods. The unintentional low gain of this detector was used to study with great precision the $\gamma$-transitions and the conversion electrons induced in the energy range of 20 to 200 keV by neutron captures in the \CaWO crystal. Thanks to the high quality of the \fifradina-\geant simulations eleven lines have been identified and a corrected decay scheme of the $^{187}$W isotope has even been proposed. Finally, the detection in coincidence of recoils in the cryogenic detector and high-energy $\gamma$-rays in the \BaF detectors outside the cryostat has been established with high significance. This world premiere validates a unique feature of the \CRAB setup to tag in time and in energy the nuclear recoils induced by neutron capture in the cryogenic detector.

Based on these promising results we propose a rich program of precision measurements with \AlO, \CaWO, Ge and Si cryogenic detectors. Beyond the calibration of nuclear recoils in the sub-keV range for all these materials, the \CRAB method provides original probes of the underlying solid-state physics. The stopping times in matter can be studied in \AlO and Si; the very low energy tungsten recoils in \CaWO provide a unique sensitivity to the energy stored in the crystal defects induced by the displacement cascades; applied on the pure nuclear recoils spectra induced by neutron capture, the simultaneous readout of the ionization and phonon channels in semiconductor detectors like Ge and Si has the potential to clarify the ionization yield for low-energy nuclear recoils, a central problem for many experiments in \CEvNS and DM communities.

The efforts are now focused on the energy resolution of the cryogenic detectors, driving the sensitivity of all measurements. With the \CaWO detectors tested so far on site, the very good performances obtained at TU-Munich in the same cryostat before its move to Vienna could not be reproduced, except for a few hours at the start of commissioning. Several improvements are underway to mitigate the electronic noise, including an improved electrical scheme of the readout circuit and the implementation of an electromagnetic shielding around the dewar. Another approach would be to further develop the coincidence technique, since it alleviates this constraint of energy resolution. The studies presented in this paper show that with the current setup the loss in statistics and the external background in the $\gamma$-detectors are the main limitations to the sensitivity. They could be partially compensated by an upgrade of the $\gamma$-detectors with better detection efficiency, more hermetic shielding against external background and larger coverage of solid angle. For longer-term operation of the cryostat, dry refrigerator technology would lighten considerably the workload at the reactor site. This would be an opportunity to optimize the entrance window of the neutron beam and further reduce the high energy background induced by the capture of scattered neutrons and increase the neutron flux. Vibration damping of the pulse tube would greatly benefit from R\&D carried out by the \RICOCHET and \NUCLEUS collaborations involved in the \CRAB project. 

After the first validation of the principle of the \CRAB method with commercial fission neutron sources \cite{PhysRevLett.130.211802,PhysRevD.108.022005}, we demonstrated here the  operation of a wet cryostat coupled to a thermal neutron beam and external $\gamma$-detectors. The proposed phasing puts the \CRAB project on the path to high-precision measurements that are of central interest for the characterization of cryogenic detectors of the neutrino and dark-matter communities, and will enable unprecedented tests of the underlying solid-state physics.

\begin{acknowledgements}
The authors are grateful for the technical and administrative support of the Atominstitut for the installation of the \CRAB experiment. We especially acknowledge the support of Roman Gergen in the design and implementation of the setup. We particularly thank Wolfgang Treimer, the Beuth Hochschule f\"ur Technik Berlin and the Helmholtz-Zentrum Berlin for ceding the 2D position sensitive detector to the Atominstitut in the context of the transfer of the instrument V12 from Berlin to Vienna. Our warmest thanks also go to Iolanda Matea and Beyhan Bastin for their precious help in retrieving the gamma detectors. We acknowledge funding by the French government under France 2030 (P2I - Graduate School Physics) under reference ANR-11-IDEX-0003, by the DFG through the SFB 1258 Excellence Cluster ORIGINS, and
by the Austrian Science Fund (FWF) through the projects
“I 5427-N CRAB” and “P 34778-N ELOISE”. 

\end{acknowledgements}


\bibliographystyle{unsrtnat}
\bibliography{bibliography}

\end{document}